\documentclass[11pt,a4]{article}

\usepackage{natbib}
\usepackage{tikz}
\usepackage{pgflibrarysnakes}
\usepackage{pgflibraryplothandlers}
\usepackage{dcolumn}%
\usepackage{amsmath}
\usepackage{amsbsy}
\usepackage{amssymb}
\usepackage{mathrsfs}
\usepackage{enumerate}
\usepackage{xcolor}
\usepackage{color}
\usepackage{nomencl}
\usepackage{fancyhdr}
\usepackage{subfigure}
\usepackage{bm}
\usetikzlibrary{plothandlers}
\usetikzlibrary{snakes}
\usepackage{epsfig}
\usepackage{float}
\usepackage{graphicx}
\usepackage{hyperref}
\hypersetup{
    colorlinks = true,
    urlcolor   = blue,
    citecolor  = black,
}
\usepackage[margin=1cm]{geometry}
\newcommand\bs{\boldsymbol{}}

\title{Sound induced by a simple impact oscillator}

\author{S. P. Narla$^1$ \and Kartik Venkatraman$^1$\thanks{kartik@iisc.ac.in}}
\date{$^1$Department of Aerospace Engineering, Indian Institute of Science, Bangalore, 560012, India}

\begin{document}

\maketitle    

\begin{abstract}
{
\noindent Acoustic radiation due to vibration and impact of a spring-mass-damper oscillator whose motion is constrained by a barrier is analyzed at a field point in a free field. Impact between the mass and the barrier is modeled using a coefficient of restitution model. Non-linear behavior of the oscillator is observed due to motion constraint. Steady state response is studied using a bifurcation diagram. For small amplitudes of oscillation, the pressure perturbation by a vibrating mass in a compressible fluid  is modeled as an acoustic dipole with its center at the equilibrium position of the mass and its axis aligned with the motion of the oscillator. The boundary condition for the acoustic domain is an acoustic free-field condition. It is observed that the unsteady acoustic pressure resulting from the impact force is a few orders of magnitude greater relative to the pressure field resulting from vibration alone before or after impact. We also analyzed the influence of coefficient of restitution, damping ratio, the ration of base excitation frequency to the natural frequency, and the ratio of the distance of the barrier to the base excitation amplitude on the acoustic radiation. Damping ratio and coefficient of restituion are shown to be the most significant paramters that affect the acoustic radiation from the vibro-impact oscillator.  
}
\end{abstract}


\makenomenclature
\nomenclature{$m$}{mass of the oscillator or mass per unit length of the beam}
\nomenclature{$k_s$}{stiffness of the spring}
\nomenclature{$c_d$}{damping coefficient}
\nomenclature{$u(t)$}{displacement of the oscillator}
\nomenclature{$u_b(t)$}{base excitation}
\nomenclature{$U_b$}{base excitation amplitude}
\nomenclature{$\omega_n$}{natural frequency of spring-mass system $\sqrt{\frac{k_s}{m}}$}
\nomenclature{$\omega_b$}{base excitation frequency}
\nomenclature{$\overline{\omega}_b$}{non-dimensional base excitation frequency $\frac{\omega_b}{\omega_n}$}
\nomenclature{$\tau$}{non-dimensional time $\omega_n t$}
\nomenclature{$\zeta$}{damping ratio $\frac{c_d}{2m\omega_n}$}
\nomenclature{$\varphi$}{phase angle between displacement response and base excitation $\tan^{-1}(\frac{c_d \omega_b}{k_s})$}
\nomenclature{$\overline{u}(\tau)$}{non-dimensional displacement of the oscillator $u(\tau)/U_b$}
\nomenclature{$\delta$}{distance between the wall and  the equilibrium position of the mass}
\nomenclature{$\overline{\delta}$}{non-dimensional distance between the wall and  the equilibrium position of the mass $\delta/U_b$} 
\nomenclature{$\tau^{\ast}$}{time of impact}
\nomenclature{$\tau^{\ast}_{+}$}{time just after impact}
\nomenclature{$\tau^{\ast}_{-}$}{time just before impact}
\nomenclature{$e$}{coefficient of restitution}
\nomenclature{$P$}{total pressure of gas}
\nomenclature{$\rho$}{density of gas}
\nomenclature{$T$}{absolute temperature in Kelvin}
\nomenclature{$R$}{gas constant in Joules/kg/K}
\nomenclature{$\Delta$}{Impact duration}
\nomenclature{$\bar{A}$}{non-dimensional acceleration amplitude of impact force $A/(U_b \omega_n^2)$}
\nomenclature{$\gamma$}{ratio of specific heats $\frac{C_p}{C_v}$}
\nomenclature{$\rho'$}{perturbation density of the fluid $\rho-\rho_0$}
\nomenclature{$p$}{acoustic or perturbation pressure $P - P_0$}
\nomenclature{$\frak{m}$}{mass injection rate}
\nomenclature{$c$}{speed of sound}
\nomenclature{$\nabla^2$}{Laplacian operator}
\nomenclature{$\theta$}{Azimuth angle}
\nomenclature{$\phi$}{Zenith angle}
\nomenclature{$\overline{r}$}{non-dimensional distance from the center of monopole or center of dipole to the field point  $r/U_b$}
\nomenclature{$\overline{p}$}{non-dimensional acoustic pressure $p/p_{0}$}
\nomenclature{$p_{0}$}{reference pressure $m\omega^2_n/ U_b$}
\nomenclature{$\overline{c}$}{non-dimensional speed of sound $c/U_b\omega_n$}
\nomenclature{$\textbf{P}_s$}{non-dimensional complex pressure amplitude of monopole}
\nomenclature{$\textbf{p}_s$}{complex pressure of monopole}
\nomenclature{$\overline{\textbf{p}}_s$}{non-dimensional complex pressure of monopole $\textbf{p}_s/p_{0}$}
\nomenclature{$\overline{\omega}$}{non-dimensional angular frequency of the spherical source $\omega/\omega_n$}
\nomenclature{$\overline{k}_a$}{non-dimensional acoustic wave number $\overline{\omega}/\overline{c}$}
\nomenclature{$\overline{\rho}$}{non-dimensional density of the medium $\rho U_b^3/m$ or  $\rho L_b^3/m$}
\nomenclature{$\overline{\textbf{R}}_s(\tau)$}{non-dimensional surface velocity of spherical source ${\textbf{R}}/U_b\omega_n$}
\nomenclature{$\overline{R}_0$}{non-dimensional velocity amplitude of the spherical source $R_0/U_b\omega_n$}
\nomenclature{$\overline{\textbf{z}}$}{non-dimensional specific acoustic impedance}
\nomenclature{$\overline{a}$}{non-dimensional radius of the spherical source $a/U_b$ or $a/L_b$}
\nomenclature{$\overline{q}_s$}{non-dimensional strength of spherical source $4\pi\overline{a}^2\overline{R}_0$}
\nomenclature{$\overline{\textbf{p}}_d$}{non-dimensional complex pressure of dipole}
\nomenclature{$\frak{U}$}{unit step function}
\nomenclature{$\frak{d}$}{Dirac-delta function}
\printnomenclature
\nobreak
\nopagebreak

\section{Introduction}
\label{sec:intro}

The analysis of the sound field generated by a flexible impacting structure is of considerable importance in noise control. Vibrating structures shows nonlinear behavior due to motion constraints.  Rattle noise inside the vehicle cabin due to seat belt retractor, gear rattle due to backlash of meshing teeth pair of the gears in gear box, tube failure due to flow induced vibration in shell and tube heat exchangers, failure of reed valve due to impact in reciprocating compressors, noise generated by punching press in punching and blanking operations, are some of the examples which involve noise generation by vibrating structures subjected to motion constraint. In all the above examples, one has to study the vibration response of the structure, the impact mechanics and the resulting acoustic field by considering the acoustic boundary conditions. 

Sound radiated from the impact of two objects has been studied in the literature as a basic problem of noise generation in mechanical structures. Akay and Bengisu \cite{Akay1983} considered the sound radiated by a thin simply supported beam due to an impact from a plexiglass ball falling at the mid-span of the beam. The acoustic radiation generated by the vibrating beam is modeled by a set of dipoles placed on the neutral axis of the beam. The acoustic field generated by the dipole is shown to be a function of the acceleration and derivative of the acceleration, also known as jerk, of the vibrating beam. Transient sound radiation from the  elastic impact of a sphere with a slab was considered by Akay and Hodgson \cite{Akay1978}. They also used Hertzian contact theory to describe the force-time history due to impact of the sphere with the slab. Sound radiation from the impacting sphere is modeled as a dipole source. The presence of the slab acts as a reflector for sound waves. This sphere-slab problem is modeled with dipole source and its image source. During impact, it was observed that peak sound pressure level is proportional to peak acceleration level. Akay and Latcha \cite{Akay1983a} investigated the acoustic field during inelastic collision of a ball with a clamped circular elastic plate. They derived an expression for displacement response of the plate using normal mode analysis. The contact force between the ball and the circular plate is modeled as a point force with a squared half-period sine wave acting at the impact point. Using the Rayleigh integral, acoustic radiation from the vibrating plate is calculated at a field point. In the calculation of  the transient sound radiation, they used time-dependent integration limits in the Rayleigh integral. This is to account for the fact that sound waves arrive early from the areas of the plate near to the field point, and also for the late arrival of waves from the distant areas of the plate to the field point.

Koss and Alfredson \cite{Koss1973} studied the transient sound field radiated by elastic collision of two spheres in free-space. The impact process was explained in terms of Hertzian contact theory. They modeled the sound field generated by each sphere during collision as that due to a finite size acoustic dipole source. The surface acceleration of the dipole source is the acceleration resulting from the impact force. They derived an expression for acoustic pressure radiated from a dipole source of arbitrary surface acceleration by using the superposition integral. Koss \cite{Koss1974} extended this procedure of determining the sound pressure to inelastic collisions between soft spheres. During impact, sound radiation from each sphere is modeled as dipole source. Acceleration-time history relation for inelastic collision was resolved into that of an elastic loading period, a  plastic-elastic loading period, and an elastic unloading period. They obtained an expression for sound field radiated from each sphere by convoluting the unit impulse pressure response with acceleration response during the inelastic collision. Total sound field at a field point in free space is obtained from the two spheres as the sum of sound field radiated by each sphere. They observed that there a peak rarefactive sound is generated at the time of elastic unloading. Their theoretically predicted sound pressure amplitudes were in good agreement with experimental values but the time of occurrence of the a peak rarefactive sound pressures were slightly different for higher impact velocities. Using similar arguments, Yufang and Zhongfang \cite{Yufang1992} considered the problem of sound radiation from two cylinders impacting side-to-side within the framework of Hertzian contact theory. They observed that the sound radiation from the impact of cylinders has directivity. They experimentally verified that it is reasonable to approximate contact acceleration with a half-sine pulse. 

Walker and  Soule \cite{Walker1996} studied the non-linear behavior of  an impact oscillator. Their study focuses on the mechanism of energy amplification due to impact and chaotic behavior of the oscillator. Tufillaro and Albano \cite{Tufillaro1986} conducted an experiment with a ball bouncing on a sinusoidally vibrating table. The collision between the ball and the table is assumed to be inelastic. As the collision is inelastic, the ball sticks to the table for small amplitudes of vibration of the table. They observed the dynamics of the ball by varying the amplitude of the vibration of the table while keeping the frequency constant. Their study concludes that the motion of the ball shows a period-doubling phenomenon before entering into the chaotic regime. Acoustic radiation due to inelastic collision of a sphere with a simply supported rectangular thin plate was considered by Troccaz et al. \cite{Troccaz2000}. Impact force-time relations are obtained by extending the Hertz's contact law taken into account the plastic deformation that occurs during inelastic collision. Using Rayleigh's integral, they calculated the acoustic pressure at a field point in free space. They observed that radiated pressure due to the initial deformation of the plate at the time of impact is significant along with the radiation due to bending waves of the plate.

In this paper, we investigate the time varying transient acoustic pressure field generated by a vibrating and impacting oscillator by coupling the  dynamics of the impacting oscillator with the dynamics of the acoustic pressure field. 

We first derive the equation of motion of the oscillator that is impacting a barrier. The impact force on the oscillator during impact is then determined using an impulse-momentum relation using a coefficient of restitution model of the impact.  We then state the in-homogenous wave equation and proceed to derive the relation for the acoustic pressure induced by a spherical source and thereafter the dipole. Later we show that the sound pressure generated by a vibrating mass can be modelled as an acoustic dipole. The discussion on the results of the numerical simulation start with the dynamics of the impact oscillator, in particular its bifurcation diagram. The bifurcation diagram is a map of the qualitative response behaviour of the nonlinear vitro-impact system. Based on the bifurcation diagram, we select few points from there to study the sound generated by the vibro-impact oscillator. The time domain response of the oscillator as a function of the coefficient of restitution as we as the damping in the oscillator is analyzed \cite{Narla2009}. The vibration as well as acoustic pressure generated by the vibration is analyzed as a function of damping ratio, excitation frequency, excitation amplitude, and coefficient restitution.

\section{Vibration model}
\label{sec:vibmod}

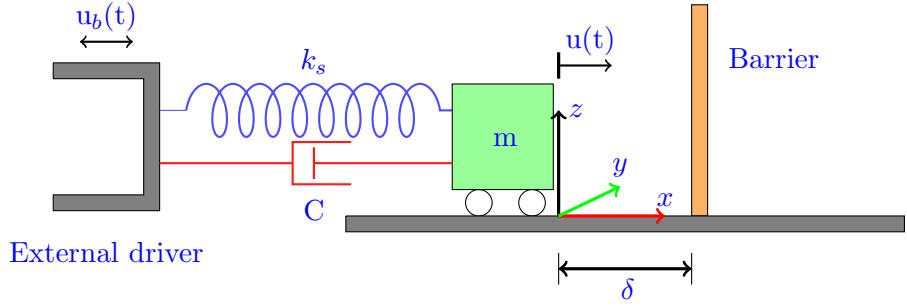
\begin{figure}
\begin{center}
\begin{tikzpicture}[scale=0.7]
\draw[snake=coil,segment amplitude=10pt,segment length=10pt,blue!70!white,thick] (0,0.5)--(5,0.5); 
\draw[fill=green!40](5,1)rectangle(6.9,-1);
\draw (5.5,-1.25)circle(0.25cm);
\draw (6.5,-1.25)circle(0.25cm);
\draw[fill=gray](-2.5,1.4)--(-2.5,1.1)--(-0.8,1.1)--(-0.8,-1.1)--(-2.5,-1.1)--(-2.5,-1.4)--(-0.5,-1.4)--(-0.5,1.4)--(-2.5,1.4);
\draw[<->,thick](-2,1.8)--(-1,1.8);
\draw[blue](-1.5,1.8)node[anchor=south]{$\textrm{u}_b(\textrm{t})$};
\draw[blue](-1.5,-1.8)node[anchor=north]{External driver};
\draw[fill=orange!60](9.8,-1.5)--(9.8,2.5)--(9.5,2.5)--(9.5,-1.5)--(9.8,-1.5);
\draw[fill=gray](3,-1.5)--(13.5,-1.5)--(13.5,-1.8)--(3,-1.8)--(3,-1.5);
\draw[->,red, very thick](7,-1.5)--(9,-1.5); 
\draw[blue](9,-1.5)node[anchor=south]{$x$}; 
\draw[->,black, very thick](7,-1.5)--(7,0.5); 
\draw[blue](7,0.5)node[anchor=west]{$z$};
\draw[->,green, very thick](7,-1.5,0)--(7,-2.1,-3);
\draw[blue](7,-2.1,-3)node[anchor=south]{$y$};
\draw[blue!70!white](0,0.5)--(-0.5,0.5);
\draw[red!90,thick](2.4,-0.5)--(5,-0.5);
\draw[red!90,thick](2.4,-0.2)--(2.4,-0.8);
\draw[red!90,thick](3.1,-0.1)--(2,-0.1)--(2,-0.9)--(3.1,-0.9);
\draw[red!90,thick](-0.5,-0.5)--(2,-0.5);
\draw[very thick](7,1.6)--(7,1.1);
\draw[->,thick](7,1.35)--(8,1.35);
\draw[<->,very thick](7,-2.5)--(9.5,-2.5);
\draw(7,-2.2)--(7,-2.8);
\draw(9.5,-2.2)--(9.5,-2.8);
\draw[blue](8.3,-2.5)node[anchor=north]{$\delta$};
\draw [blue](7.6,1.35)node[anchor=south]{u(t)};
\draw[blue] (10,1.5) node [anchor=west]{Barrier};
\draw[blue] (2.4,1)node[anchor=south]{$k_s$};
\draw[blue] (2.4,-1.0)node[anchor=north]{C};
\draw[blue] (6,0.25)node[anchor=north]{m};
\end{tikzpicture}
\vspace{0.6cm}
\caption{Impact oscillator}
\label{f:hull} 
\end{center} 
\end{figure}

Figure~\ref{f:hull} shows a mass connected to an external driver through spring and damper. The equation of motion for the oscillator is given by the relation 

\begin{equation}
m\ddot{u}(t)+c_d\dot{u}(t)+k_su(t)=k_s u_b(t)+c_d\dot{u}_b(t),
\end{equation}

where the variables and parameters are as defined in the Nomenclature. The spring-mass system is excited by the external driver whose displacement is time varying and is defined as $u_b(t)=U_b\cos(\omega_b t)$. Then the above equation becomes

\begin{equation}
\label{sdofmodel}
m\ddot{u}(t)+c_d\dot{u}(t)+k_s u(t)=U_b\sqrt{{k_s}^2+(c_d\,\omega_b)^2}\cos(\omega_b t+\varphi).
\end{equation}

Equation~(\ref{sdofmodel}) can be expressed in non-dimensional form as

\begin{equation}
\label{e:presen}
\overline{u}{''}(\tau)+2\zeta \overline{u}{'}(\tau)+\overline{u}(\tau)=\sqrt{1+(2\zeta\overline{\omega}_b)^2}\cos(\overline{\omega}_b \,\tau +\varphi),
\end{equation}

with initial conditions $\left. \overline{u}\right|_{\tau=0}=\frac{u_0}{U_b}$ and $\left. {\overline{u}}{'}\right|_{\tau=0}=\frac{\dot{u}_0}{\omega_n U_b}$; $u_0$ and $\dot{u}_0$ are the initial displacement and initial velocity of the mass. $U_b$ is the base excitation amplitude, $\overline{\omega}_b=\frac{\omega_b}{\omega_n}$, $\zeta=\frac{c_d}{2m{\omega}_n}$ and $\varphi=\tan^{-1}(2\zeta\overline{\omega}_b)$.  Prime $(')$ is used here to denote that differentiation is with respect to $\tau=\omega_n t$.

We assume that initially no barrier is present. After the mass completes certain number of cycles of oscillations about its equilibrium position, the barrier is placed at a distance $\overline{\delta}$ from the unextended static equilibrium position of the spring stiffness of the oscillator. The collision between the mass and the barrier is modeled using the coefficient of restitution $e$ such that $0\leq e \leq 1 $. The equation of motion of the oscillator with this motion constraint is 

\begin{equation}
\label{e:presen}
\overline{u}{''}(\tau)+2\zeta \overline{u}{'}(\tau)+\overline{u}(\tau)=\sqrt{1+(2\zeta\overline{\omega}_b)^2}\cos(\overline{\omega}_b \,\tau +\varphi):  \quad \text{when }\overline{u}(\tau)\leq \overline{\delta},
\end{equation}

or equivalently the mass is not impacting the barrier. At the moment of impact there is displacement continuity and velocity reversal given by the restitution law. That is

\begin{equation}
\label{e:velocity reversal}
 \overline{u}(\tau^{\ast}_{+})=\overline{u}(\tau^{\ast}_{-})=\overline{\delta}, \quad \overline{u}{'}(\tau^{\ast}_{+}) = -e\,\overline{u}{'}(\tau^{\ast}_{-}).
\end{equation}

In the above set of equations, $\tau^{\ast}_{-}$ is the time just before impact,  $\tau^{\ast}_{+}$ is the time just after impact,  $\overline{\delta}$ is the non-dimensional distance between the barrier and the equilibrium position of the spring when it is unforced.

\section{Impact model}
\label{sec:impmod}

From Equation~(\ref{e:velocity reversal}), the acceleration of the oscillator mass is infinite at the time of impact due to instantaneous reversal of velocity. Since this is neither practically feasible nor computationally realisable, for the purpose of numerical computation, we assume that the impact will take place during a finite time duration $\Delta =\frac{1}{100\, \overline{\omega}_b}$. We also assume a sinusoidal reaction force acting on the mass during impact. As a result of the sinusoidal force, the acceleration during impact is $\overline{a}_r(\tau)=\bar{A}\sin\bigl(\frac{\pi \tau}{\Delta} \bigr)$. Using the impulse-momentum relation, the change in velocity after and before impact is given by

\begin{equation}
\label{eq:momentu1}
\quad \overline{u}{'}(\tau^{\ast}_{+})-\overline{u}{'}(\tau^{\ast}_{-} )=\int_{{\tau^{\ast}_{-}}}^{{\tau^{\ast}_{+}}} \bar{A}\sin\bigl[\frac{\pi}{\Delta}(\epsilon-\tau^{\ast}_{-}) \bigr]d \epsilon,
\end{equation}
 
 where $\epsilon$ is a dummy variable of integration,  $\tau^{\ast}_{+}=\tau^{\ast}_{-}+\Delta$ and  $\overline{u}{'}(\tau^{\ast}_{+})=-e\overline{u}{'}(\tau^{\ast}_{-})$. Solving for $\bar{A}$ using  Equation~(\ref{eq:momentu1}), we get
 
\begin{equation}
\label{eq:impact force acceleration amplitude}
\bar{A}=-\frac{\pi\,\overline{u}{'}(\tau^{\ast}_{-})(1+e)}{2\Delta}.
\end{equation}
 
Therefore, the expressions for acceleration and jerk---the rate of change of acceleration---during impact are 
 
\begin{equation}
\label{imaccamp}
\overline{a}_r(\tau) = \bar{A}\sin\bigl( \frac{\pi \tau}{\Delta} \bigr) \\
\end{equation}

and

\begin{equation}
\label{imjerkamp}
\overline{a}_r{'}(\tau) = \frac{\pi}{\Delta}\bar{A}\cos\bigl( \frac{\pi \tau}{\Delta} \bigr)
\end{equation} 

respectively.

\section{Acoustic model}
\label{sec:acomod}

We present here an outline to the modeling of the acoustic radiation by a vibrating mass. We begin with the main steps in the derivation leading to the wave equation with source terms. These source terms could be either due to mass injection or volume perturbation in the medium, or application of force. Finally we deal with the dipole model to represent the forced pressure perturbation of the wave equation.

\subsection{Acoustic wave equation}
\label{sec:acowaveq}

The derivation here essentially follows that of \citep[Chap.3]{Fahy2000}.

We assume that the fluid medium behaves as an ideal gas would, and that the thermodynamic process during small amplitude pressure-density perturbation at audio frequencies in an ideal gas is isentropic.

The ideal gas relation is given by

\begin{equation}
\label{eq:waveeq}
\frac{P}{\rho} = R T.
\end{equation}

The thermodynamic process during small amplitude pressure-density perturbation at audio frequencies in an ideal gas is adiabatic - there is no heat addition or deletion. The adiabatic relation is

\begin{equation}
\label{eq:adiabat2}
\frac{P}{\rho^\gamma} = \text{constant}.
\end{equation}

Differentiating 

\begin{equation}
\label{eq:adiabat2}
\Bigl(\frac{dP}{d\rho}\Bigr)_0 = \gamma\frac{P_0}{\rho_0} = \gamma R T_0.
\end{equation}

The speed of sound is defined as 

\begin{equation}
\label{eq:soundspeed}
c^2 \triangleq \Bigl(\frac{dP}{d\rho}\Bigr)_0 = \gamma\frac{P_0}{\rho_0} = \gamma R T_0.
\end{equation}

The adiabatic relation also gives a relation between the pressure and density perturbations due to an acoustic disturbance through the medium as

\begin{equation}
\label{eq:adiabat3}
p = \frac{\gamma P_0}{\rho_o} \rho'.
\end{equation}

The perturbation density is defined as $\rho '\triangleq \rho - \rho_0$ and the pressure perturbation is defined as  $p \triangleq P-P_0$. 

The mass conservation principle applied to a control volume leads to the continuity equation with a source term

\begin{equation}
\label{eq:conteq}
\frac{\partial \rho}{\partial t} + \nabla\cdot(\rho\bs{q}) = \frac{\partial \frak{m}}{\partial t}.
\end{equation}

Expressed in terms of the perturbation density $\rho$, and after ignoring second order terms in the perturbed variables, the continuity equation can be represented as 

\begin{equation}
\label{eq:conteq2}
\frac{\partial \rho'}{\partial t} + \rho_0\nabla\cdot\bs{q} = \frac{\partial \frak{m}}{\partial t}.
\end{equation}

The density perturbation can be related to the change in volume $\delta V$ of the gas as

\begin{equation}
\label{eq:dens-volume}
\frac{\rho'}{\rho_0} = \frac{\delta V}{V_0}.
\end{equation}

One can go further, and express Equation~(\ref{eq:conteq2}) in terms of the pressure perturbation $p$ rather than the density perturbation using the relation $p=c^2\rho'$

\begin{equation}
\label{eq:conteq3}
\frac{1}{c^2}\frac{\partial p}{\partial t} + \rho_0\nabla\cdot\bs{q} = \frac{\partial \frak{m}}{\partial t}.
\end{equation}

The linear momentum conservation equation with external excitation term is given by

\begin{equation}
\label{eq:linmom1}
\rho_0\frac{\partial\bs{u}}{\partial t} = -\nabla P + \bs{f}.
\end{equation}

In terms of the perturbed quantities, the linear momentum equation can be represented as

\begin{equation}
\label{eq:linmom2}
\rho_0\frac{\partial\bs{u}}{\partial t} = -\nabla p + \bs{f}.
\end{equation}

Now taking the time derivative of Equation~(\ref{eq:conteq3}) and the gradient of Equation~(\ref{eq:linmom2}), and eliminating the cross-derivative term, we get the wave equation in terms of the perturbation pressure

\begin{equation}
\label{eq:waveeq}
\nabla^2 p - \frac{1}{c^2}\frac{\partial^2 p}{\partial t^2} = \nabla\cdot\bs{f} - \frac{\partial^2 \frak{m}}{\partial t^2}.
\end{equation}

In case there are no sources and sinks, nor external forces, the right-hand-side will be zero. In what follows, we will not pursue modeling the mass injection term $\partial^2 \frak{m}/\partial t^2$. Rather, we would be interested in the external excitation term since our interest is in acoustic perturbations excited by a vibrating and impacting rigid mass in a compressible inviscid fluid medium.

\subsection{Pulsating sphere}
\label{sec:pulsphmod}

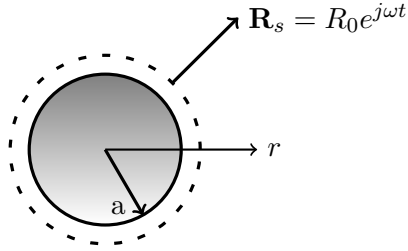
\begin{figure}[h]
\begin{center}
\begin{tikzpicture}[scale=0.5]
\shadedraw[very thick] (0,0) circle (2cm);
\draw[loosely dashed,very thick] (0,0) circle (2.5cm);
\draw[->,very thick]  (0,0)--(1,-1.732);
\draw [->,thick] (0,0)--(4,0);
\draw[->,very thick] (1.7677,1.7677)--(3.5,3.5);
\draw (0.8,-1.532) node[anchor=east]{a};
\draw (4,0) node[anchor=west]{$r$};
\draw (3.5,3.5) node[anchor=west]{$\textbf{R}_s=R_0e^{j\omega t}$};
\end{tikzpicture}
\vspace{0.6cm}
\caption{Pulsating sphere}
\label{f:bhavitha1} 
\end{center}
\end{figure}

A pulsating sphere is a sphere whose radius varies sinusoidally with time as shown in Figure~\ref{f:bhavitha1}.

Let the radius of the pulsating sphere be $a$. The radial velocity $R_s$ of any point on its surface is given by 

\begin{equation}
R_s(t)=R_0\cos{\omega t}.
\label{e:fattigue1}
\end{equation}

The acoustic wave equation is represented as

\begin{equation}
\label{eq:lalitha1}
\frac{\partial^2 p}{\partial t^2}=c^2 \nabla^2 p,
\end{equation}

where $c$ is the velocity of sound in a medium. In spherical coordinates the Laplacian operator is expressed as

\begin{equation}
\nabla^2=\frac{\partial^2}{\partial r^2}+\frac{2}{r}\frac{\partial}{\partial r}+\frac{1}{r^2\sin\phi}\frac{\partial}{\partial \phi}\left(\sin\phi\frac{\partial}{\partial \phi}\right)+\frac{1}{r^2\sin^2\phi}\frac{\partial^2}{\partial \theta^2}. 
\end{equation}

If the waves have spherical symmetry, that is, if the acoustic pressure is a function of radial distance and time but not of the angular coordinates $\theta$ and $\phi$, then the wave Equation~(\ref{eq:lalitha1}), takes the form 

\begin{equation}
\label{eq:lali1}
\frac{\partial^2 }{\partial t^2}\left(rp(r,t) \right)=c^2\frac{\partial^2}{\partial r^2}\left(rp(r,t) \right).
\end{equation}

Re-scaling the radial distance, time, pressure, frequency, wave number, and the speed of sound in air, then Equation~(\ref{eq:lali1}) is represented as 

\begin{equation}
\label{eq:appoo1}
\frac{\partial^2 (\overline{r}\,\overline{p})}{\partial \tau^2}=\overline{c}^2\frac{\partial^2(\overline{r}\,\overline{p})}{\partial \overline{r}^2}.
\end{equation}

The solution of the wave equation Equation~(\ref{eq:appoo1}) are diverging spherical harmonic waves, which in complex form is expressed as 

\begin{equation}
\overline{\textbf{p}}_s=\frac{\textbf{P}_s}{\overline{r}}e^{j(\overline{\omega}\,\tau-\overline{k}_a\,\overline{r})}.
\end{equation}

Note that the actual pressure $\overline{p}_s = Re(\overline{\textbf{p}}_s)$.

The radial particle velocity for spherical harmonic waves in complex form is

\begin{equation}
\overline{\textbf{R}}_s=-\frac{1}{j\,\overline{\omega}\,\overline{\rho}} \frac{\partial \overline{\textbf{p}}_s}{\partial \overline{r}}=\left(\frac{1}{\overline{r}}+j\overline{k}_a\right)\frac{\overline{\textbf{p}}_s}{j\,\overline{\omega}\,\overline{\rho}}.
\end{equation}

The specific acoustic impedance is defined as 

\begin{equation}
\label{eq:acous1}
\overline{\textbf{z}}=\frac{\overline{\textbf{p}}_s}{\overline{\textbf{R}}_s}=\frac{j\,\overline{\omega }\,\overline{\rho}}{\left(\frac{1}{\overline{r}}+j\,\overline{k}_a\right)}.
\end{equation}

The complex form of the non-dimensional spherical source surface velocity Equation~(\ref{e:fattigue1}) is 

\begin{equation}
\overline{\textbf{R}}_s(\tau)=\overline{R}_0 e^{j\,\overline{\omega}\,\tau}.
\end{equation}

The fluid medium surrounding the sphere must remain in contact with the surface of the sphere all the time, so the particles surrounding the sphere have the same velocity as the surface velocity of the sphere. Therefore at $\overline{r}=\overline{a}$

\begin{equation}
\label{eq:radialvel}
\overline{R}_0 e^{j\,\overline{\omega}\,\tau} = \frac{\overline{\textbf{p}}_s}{\overline{\textbf{R}}_s} = \frac{\textbf{P}_s}{\overline{a}\, \overline{\textbf{z}}_{\overline{r}=\,\overline{a}}}\,e^{j(\overline{\omega}\, \tau-\overline{k}_a\,\overline{a})}
\end{equation}

\noindent where $\overline{\textbf{z}}_ {\overline{r}=\,\overline{a}}$ is the specific acoustic impedance of a spherical wave evaluated at $\overline{r}=\overline{a}$. Substituting Equation~(\ref{eq:acous1}) in Equation~(\ref{eq:radialvel}) we get

\begin{equation}
\textbf{P}_s= \overline{a}\overline{R}_0\,\overline{\textbf{z}}_{\overline{r}=\overline{a}}\,e^{j\,\overline{k}_a\,\overline{a}} =\overline{a} \overline{R}_0\frac{j\,\overline{\omega}\,\overline{\rho}}{\left(\frac{1}{\overline{a}}+j\,\overline{k}_a\right)}(\cos{\overline{k}_a\,\overline{a}}+j\,\sin{\overline{k}_a\,\overline{a}}).
\end{equation}

If the radius of the pulsating sphere is so small that $\overline{k}_a\,\overline{a}<<1$ at all operating frequencies

\begin{equation}
\textbf{P}_s\approx j\,\overline{a}^2\,\overline{R}_0\,\overline{\omega}\,\overline{\rho}.
\end{equation}

The equation for the acoustic pressure is therefore

\begin{equation}
\overline{\textbf{p}}_s=\frac{j\,\overline{a}^2\,\overline{R}_0\,\overline{\omega}\,\overline{\rho}}{\overline{r}} e^{j(\overline{\omega}\,\tau-\overline{k}_a\,\overline{r})}.
\end{equation}

The real part of this expression represents the actual pressure in the wave, that is

\begin{equation}
\overline{p}_s=\frac{-\, \overline{a}^2\,\overline{R}_0\,\overline{\omega}\,\overline{\rho}}{\overline{r}}\sin(\overline{\omega}\,\tau-\overline{k}_a\,\overline{r}).
\end{equation}

The strength of the spherical source $\overline{q}_s$ is defined as the product of its surface area and velocity amplitude, that is $\overline{q}_s=4\pi\overline{a}^2\,\overline{R}_0$. The equation for the acoustic pressure is then represented as 

\begin{equation}
\overline{\textbf{p}}_s=\frac{j\,\overline{\omega}\,\overline{\rho}\,\overline{q}_s}{4\pi \overline{r}} e^{j(\overline{\omega}\,\tau-\overline{k}_a\,\overline{r})}.
\end{equation}

\subsection{Dipole source}
\label{sec:dipsou}

\begin{figure}
\begin{center}
\begin{tikzpicture}[scale=1][font=\small]
\draw[red,thick,->](4,1,4.5)--(4,3.5,4.5);
\draw[yellow!80!blue,thick,->](4,1,4.5)--(4,0.3,1.5);
\draw[dashed](4,1,4.5)--(6.0,0,0.5);
\draw[dashed](2,1,4.5)--(6,3,0.5);
\draw[dashed](2,1,4.5)--(4,1,4.5);
\draw[ball color=orange] (2,1,4.5) circle (.1cm); 
\draw[blue,thick,->](6,1,4.5)--(8,1,4.5);
\draw[<->](4,0.4,4.5)--(6,0.4,4.5);
\draw[<->](2,0.4,4.5)--(4,0.4,4.5);
\draw(4,0.5,4.5)--(4,0.3,4.5);
\draw(2,0.5,4.5)--(2,0.3,4.5);
\draw(6,0.5,4.5)--(6,0.3,4.5);
\draw(2,0.5,4.5)--(2,0.3,4.5);
\draw(5,0.4,4.5)node[anchor=north]{$\frac{l}{2}$};
\draw(3,0.4,4.5)node[anchor=north]{$\frac{l}{2}$};
\draw(6.2,1,4.5)node[anchor=north]{$+q_s$};
\draw(1.8,1,4.5)node[anchor=north]{$-q_s$};
\draw[blue](4.6,1.7,1.5)node[anchor=south]{$r$};
\draw[blue](5.5,2,1.5)node[anchor=north]{$r_1$};
\draw[blue](3.5,2,1.5)node[anchor=north]{$r_2$};
\draw[black](4,3.5,4.5)node[anchor=south]{z};
\draw[black](8,1,4.5)node[anchor=west]{x};
\draw[black](4,0.3,1.5)node[anchor=south]{y};
\draw[dashed](4,1,4.5)--(6.0,3.0,0.5);
\draw[dashed](6,1,4.5)--(6.0,3.0,0.5);
\draw[dashed](6.0,3.0,0.5)--(6.0,0,0.5);
\draw[blue](6.0,0,0.5)node[anchor=west]{$\textrm{s}_p(r\sin{\phi},\theta)$};
\draw[dashed](6,1,4.5)--(6.0,0,0.5);
\draw[blue,thick](4,1,4.5)--(6,1,4.5);
\draw[blue](6.0,3.0,0.5)node[anchor=south]{$S(r,\theta,\phi)$}; 
\draw[red!60!blue,->](5.2,1,4.5)arc(0:19:0.6);
\draw[blue](5.3,1.2,4.7)node[anchor=west]{$\theta$};
\draw[red!20!blue,->](4,1.5,4.5)arc(70:30:0.5);
\draw[blue](4.1,1.6,4.5)node[anchor=west]{$\phi$};
\shade[ball color=green] (6,1,4.5) circle (.1cm);
\end{tikzpicture}
\vspace{0.6cm}
\caption{Acoustic doublet}
\label{f:rats1} 
\end{center}
\end{figure}
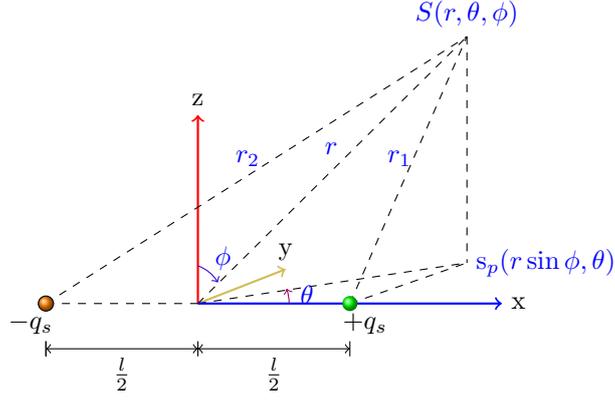

Consider two small spherical sources of equal strength $\overline{q}_s$ radiating simultaneously, but pulsating with a relative phase difference of $180^\circ$ as shown in Figure~\ref{f:rats1}. The acoustic pressure produced by a small spherical source of radius $\overline{a}$ at low frequencies for which $\overline{k}_a\,\overline{a}\ll 1$ is given by 

\begin{equation}
\overline{\textbf{p}}_s=\frac{j\,\overline{\omega}\,\overline{\rho}\,\overline{q}_s}{4\pi \overline{r}} e^{j(\overline{\omega}\,\tau-\overline{k}_a\,\overline{r})}.
\end{equation}

The acoustic pressure $\overline{\textbf{p}}_d$ of the  dipole at a field point $S(\overline{r},\theta,\phi)$ due to radiation from both sources, as discussed in  \cite{Kinsler2000}, is

\begin{equation}
\label{eq:pradeep1}
\overline{\textbf{p}}_d = \displaystyle\frac{j\,\overline{\omega}\,\overline{\rho}\,\overline{q}_s}{4\pi}e^{j\,\overline{\omega}\,\tau}
\left(\frac{e^{-j\,\overline{k}_a\,\overline{r}_1}}{\overline{r}_1}-\frac{e^{-j\,\overline{k}_a\,\overline{r}_2}}{\overline{r}_2}\right).
\end{equation}

The distance of the field point from the two spherical monopoles constituting the dipole is

\begin{eqnarray}
\label{eq:distdp}
{\overline{r}_1}^2 &=&\overline{r}^2+\left(\frac{\overline{l}}{2}\right)^2-\overline{r}\,\overline{l}\cos\theta \sin\phi\\
{\overline{r}_2}^2 &=&\overline{r}^2+\left(\frac{\overline{l}}{2}\right)^2+\overline{r}\,\overline{l}\cos\theta \sin\phi.
\end{eqnarray}

Since $\overline{r}$ is large as compared to $\overline{l}$, the above equations can be simplified to

\begin{eqnarray}
\label{eq:distdp2}
\overline{r}_1\approx & 
&\overline{r}\left(1-\frac{\overline{l}}{\overline{r}}\cos\theta\sin\phi\right)^\frac{1}{2}\approx\overline{r}\left(1-\frac{\overline{l}\cos\theta\sin\phi}{2\overline{r}}\right) \\ 
\overline{r}_2\approx & 
&\overline{r}\left(1+\frac{\overline{l}}{\overline{r}}\cos\theta\sin\phi\right)^\frac{1}{2}\approx\overline{r}\left(1+\frac{\overline{l}\cos\theta\sin\phi}{2\overline{r}}\right).
\end{eqnarray}

In the case where the distance between the two sources is so small that $\overline{l}\ll \overline{r}$ and $\overline{k}_a\,\overline{l}\ll1$,  Equation~(\ref{eq:pradeep1}) is simplified to give the pressure field of an acoustic doublet as

\begin{equation}
\overline{\textbf{p}}_d=\frac{j\,\overline{\omega}\,\overline{\rho}\,\overline{q}_s}{4\pi \overline{r}} e^{j(\overline{\omega}\,\tau-\overline{k}_a\,\overline{r})}\left[\frac{2j\sin[\frac{\overline{k}_a\,\overline{l}\cos\theta\sin\phi}{2}]+\frac{\overline{l}\cos\theta\sin\phi}{\overline{r}}\cos[\frac{\overline{k}_a\,\overline{l}\cos \theta\sin\phi}{2}]}{1-\frac{{\overline{l}}^2\cos^2\theta\sin^2\phi}{4{\overline{r}}^2}}\right].
\label{e:fot1}
\end{equation}

For $\overline{k}_a\,\overline{l}\ll1$ we use the approximation

\begin{eqnarray}
\sin\left(\frac{\overline{k}_a\,\bar{l}\cos\theta\sin\phi}{2}\right)\approx & \frac{\overline{k}_a\,\overline{l}\cos\theta\sin\phi}{2} \\
\cos\left(\frac{\overline{k}_a\,\overline{l}\cos \theta\sin\phi}{2}\right)\approx & 1.
\end{eqnarray}

For $\overline{l}\ll\overline{r}$, the term $ {\overline{l}}^2\cos^2\theta\sin^2\phi/4\overline{r}^2$ in the denominator of Equation~(\ref{e:fot1}) is neglected, and Equation~(\ref{e:fot1}) becomes 

\begin{equation}
\label{eq:dipolefreq}
\overline{\textbf{p}}_d(\overline{r},\theta,\phi,\tau)=\frac{j\,\overline{\omega}\,\overline{\rho}\,\overline{q}_s}{4\pi \overline{r}}e^{j(\overline{\omega}\,\tau-\overline{k}_a\,\overline{r})}\left[j\,\overline{k}_a\,\overline{l}+\frac{\overline{l}}{\overline{r}}\right]\cos\theta\sin\phi.
\end{equation} 

In terms of the surface velocity of the monopole constituting the dipole

\begin{equation}
\label{e:sit1}
\overline{\textbf{p}}_d(\overline{r},\theta,\phi,\tau)=\frac{j\,\overline{l}\,\overline{\omega}\,\overline{\rho}\,\overline{a}^2}{\overline{r}}\overline{\textbf{R}}_s(\tau)e^{-j\,\overline{k}_a\,\overline{r}}\left[j\overline{k}_a+\frac{1}{\overline{r}}\right]\cos\theta\sin\phi.
\end{equation}

For a dipole whose surface vibrates with an arbitrary velocity $\overline{\textbf{R}}_s(\tau)$, the Fourier transform of $\overline{\textbf{R}}_s(\tau)$ is 

\begin{equation}
\overline{\textbf{R}}_s(\overline{\omega})=\int_{-\infty}^{+\infty} \overline{\textbf{R}}_s(\tau) e^{-j\,\overline{\omega}\,\tau}\mathrm{d}{\tau}.
\end{equation}

By taking the Fourier transform on both sides of Equation~(\ref{e:sit1}) we get 

\begin{equation}
\label{dipolepresfreq}
\overline{\textbf{p}}_d(\overline{r},\theta,\phi,\overline{\omega})=\frac{\overline{l}\,\overline{\rho}\,\overline{a}^2}{\overline{r}}\overline{\textbf{R}}_s(\overline{\omega})e^{-j\,\overline{k}_a\,\overline{r}}\left[\frac{1}{\bar{c}}(j\,\overline{\omega})^2+\frac{1}{\overline{r}}(j\,\overline{\omega})\right]\cos\theta\sin\phi.
\end{equation}

The pressure-time history can be obtained from the above equation by inverse Fourier transform. That is

\begin{equation}
\label{eq:prad1}
\overline{\textbf{p}}_d(\overline{r},\theta,\phi,\tau)=\frac{\overline{l}\,\overline{\rho}\,\overline{a}^2}{\overline{r}}\cos\theta\sin\phi\, \frac{1}{2\pi}\int_{-\infty}^{+\infty} \left[\frac{1}{\overline{c}}(j\,\overline{\omega})^2+\frac{1}{\overline{r}}(j\overline{\omega})\right] \overline{\textbf{R}}_s(\overline{\omega})e^{-j\,\frac{\overline{\omega}\,\overline{r}}{\overline{c}}}e^{j\,\overline{\omega}\,\tau}\,\mathrm{d}\overline{\omega}.
\end{equation}

By using the time-shift property of a Fourier transform, Equation~(\ref{eq:prad1}) can be written as

\begin{equation}
\label{eq:prthd1}
\overline{\textbf{p}}_d(\overline{r},\theta,\phi,\tau+\frac{\overline{r}}{\overline{c}})=\frac{\overline{l}\overline{\rho}\,\overline{a}^2}{\overline{r}}\cos\theta\sin\phi\,
\frac{1}{2\pi}\int_{-\infty}^{+\infty} \left[\frac{1}{\overline{c}}(j\overline{\omega})^2+\frac{1}{\overline{r}}(j\overline{\omega})\right] \overline{\textbf{R}}_s(\overline{\omega})e^{j\,\overline{\omega}\,\tau}\,\mathrm{d}\overline{\omega}.
\end{equation}

Finally using the differentiation property of the Fourier transform, Equation~(\ref{eq:prthd1}) takes the form

\begin{equation}
\label{eq:prthde1}
\overline{\textbf{p}}_d(\overline{r},\theta,\phi,\tau+\frac{\overline{r}}{\overline{c}})=\frac{\overline{l}\,\overline{\rho}\,\overline{a}^2}{\overline{r}}\cos\theta\sin\phi\left[\frac{1}{\overline{c}}\frac{\partial^2 \overline{\textbf{R}}_s}{\partial \tau^2}(\tau)+\frac{1}{\overline{r}}\frac{\partial \overline{\textbf{R}}_s}{\partial \tau}(\tau)\right].
\end{equation}

Time-shifting once again, Equation~(\ref{eq:prthde1}) can be represented as 

\begin{equation}
\label{e:tpressure1}
\overline{\textbf{p}}_d(\overline{r},\theta,\phi,\tau)=\frac{\overline{l}\overline{\rho}\,\overline{a}^2}{\overline{r}}\cos\theta\sin\phi\left[\frac{1}{\overline{c}}\frac{\partial^2 \overline{\textbf{R}}_s}{\partial \tau^2}(\tau-\frac{\overline{r}}{\overline{c}})+\frac{1}{\overline{r}}\frac{\partial \overline{\textbf{R}}_s}{\partial \tau}(\tau-\frac{\overline{r}}{\overline{c}})\right].
\end{equation}

For pressure field in $xy$ plane, we have $\phi=\pi/2$ and the above Equation~(\ref{e:tpressure1}) becomes 

\begin{equation}
\label{eq:pressure filed in xy plane}
\overline{\textbf{p}}_d(\overline{r},\theta,\tau)=\frac{\overline{l}\overline{\rho}\,\overline{a}^2}{\overline{r}}\cos\theta\left[\frac{1}{\overline{c}}\frac{\partial^2 \overline{\textbf{R}}_s}{\partial \tau^2}(\tau-\frac{\overline{r}}{\overline{c}})+\frac{1}{\overline{r}}\frac{\partial \overline{\textbf{R}}_s}{\partial \tau}(\tau-\frac{\overline{r}}{\overline{c}})\right].
\end{equation}

In general the real part of the above Equation~(\ref{eq:pressure filed in xy plane}) gives the acoustic pressure  at the field point. In the $x-y$ coordinate system, the acoustic pressure at the field point $S(\bar{x}_f,\bar{y}_f,)$  is given by 

\begin{equation}
\label{eq:dipole pressure in xy coordinate system}
\overline{p}_d(\overline{x}_f, \overline{y}_f,\tau+\frac{\overline{r}}{\overline{c}})=\frac{\overline{l}\overline{\rho}\,\overline{a}^2}{\overline{r}}\left[\frac{1}{\overline{c}}\overline{u}{'''}(\tau)+\frac{1}{\overline{r}}\overline{u}{''}(\tau)\right]\frac{\overline{x}_f}{\overline{r}}.
\end{equation}

where $\overline{r}=[\overline{x}_f^2+\overline{y}_f^2]^{\frac{1}{2}}$.

\subsection{Oscillating sphere in a compressible fluid}
\label{sec:oscsphcomfld}

The discussion here follows the arguments put forth in \cite[pp.98-103,pp.112-118]{Fahy2000}. Let us assume that a moving boundary has dimensions transverse to its direction of motion small compared with the wavelength of the acoustic disturbance induced. Therefore the volume displaced by it is negligible. However, linear momentum fluctuations are induced in the fluid, and thereby a pressure disturbance is caused. The pressure over a surface area of the fluid is the force in a direction perpendicular to the surface. From Equation~(\ref{eq:waveeq}), the external force on the fluid appears as a divergence term. Note that $\bs{f}$ is a force per unit volume. If this force acts on a disk of unit area $\delta S$ and thickness $n$, the force can be represented as a spatial pulse with support $n_1\le z\le n_2$ 

\begin{equation}
\label{eq:forcepulse}
\begin{aligned}
\bs{f} &= f\delta S\bigl(\frak{U}(n-n_1) - \frak{U}(n-n_2)\bigr)\hat{\bs{n}} \\
\nabla\cdot\bs{f} &= f\delta S\bigl(\frak{d}(n-n_1) - \frak{d}(n-n_2)\bigr)\hat{\bs{n}}.
\end{aligned}
\end{equation}

Since these pair of concentrated forces are the result of a pressure disturbance, and are of opposite sign, they can be replaced by an acoustic dipole, Equations~(\ref{eq:dipolefreq}) or (\ref{e:sit1}).

\begin{figure}
\centering
\includegraphics[scale=0.4]{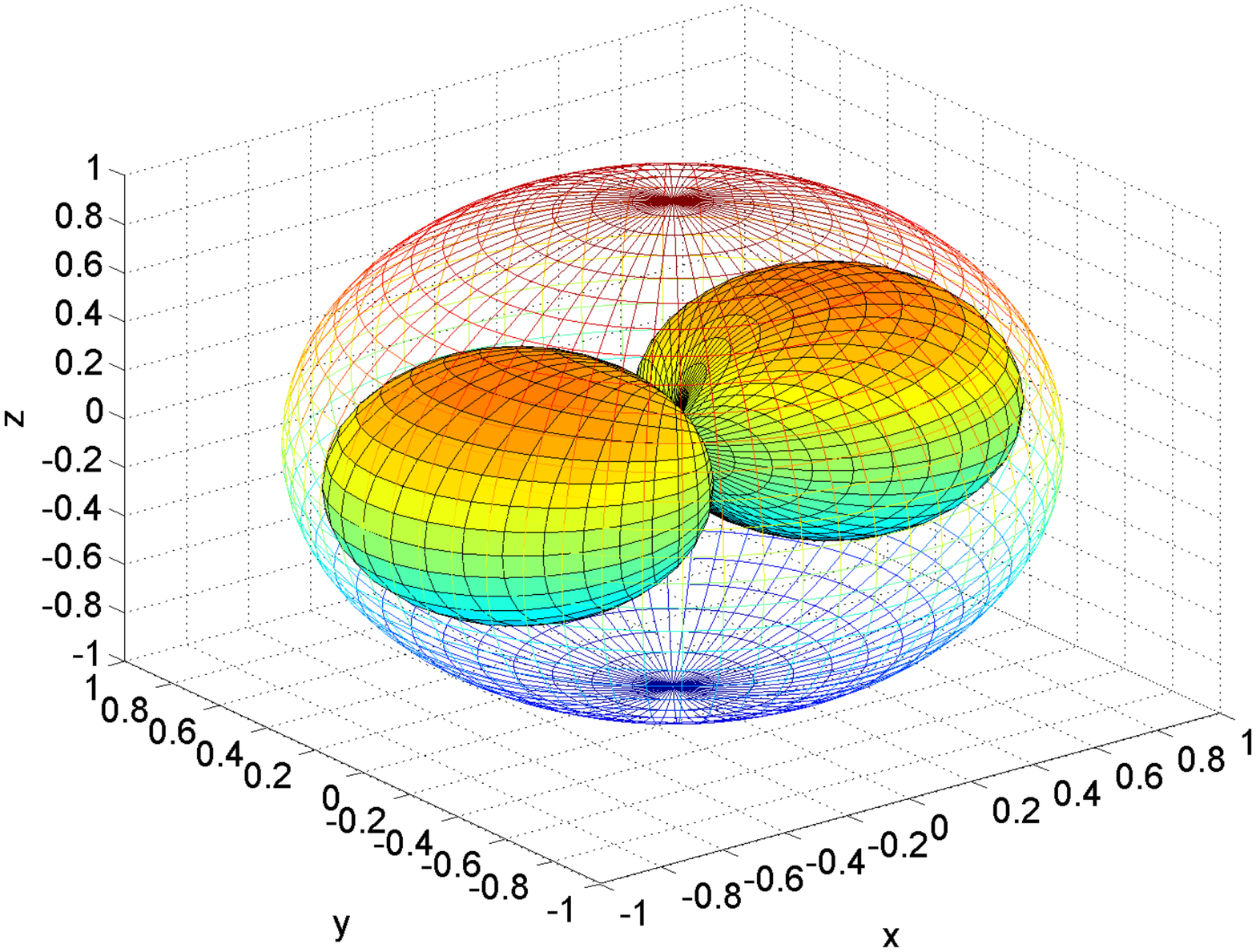}
\caption{Pressure contours of a dipole aligned with $x$ axis. $\overline{\textbf{p}}_d(\theta,\phi)=\overline{A} \cos\theta \sin\phi$; $\overline{A}=1$}
\label{f:dipolepress}
\end{figure}

An oscillating sphere or for that matter a slender cylinder whose dimensions transverse to its axis is relatively small causes no net volume changes in the fluid. However, as it displaces the compressible fluid, a net force acts on it. By the law of reaction, the oscillating rigid body exerts a net force on the fluid. This force creates pressure perturbations, the nature of which is such that it can be modeled as an acoustic dipole with the axis of the dipole along the axis of the cylinder. The dipole pressure perturbation as a function of oscillation frequency of the vibrating sphere is given by Equation~(\ref{eq:dipolefreq}). Figure~\ref{f:dipolepress} shows the dipole pressure distribution when the dipole axis is aligned in the $x$ direction. If one were to draw an arbitrary circle in this plane enclosing the origin of the dipole axis, the net force in the $y$ direction will be zero. Similarly, if one were to project the pressure distribution due to the dipole on to the $x-z$ plane, then the net force in the $z$ direction will be zero. There will be a non-zero force only in the $x$ direction.In order to show that this is indeed the case, we refer to Equation~(\ref{e:tpressure1}). At a given instant of time $\tau$ and at given radial distance $\overline{r}$, Equation~(\ref{eq:dipolefreq}) is of the form

\begin{equation}
\label{eq:tpressuredi}
\overline{\textbf{p}}_d(\theta,\phi)=\overline{A} \cos\theta \sin\phi,
\end{equation}

where $\overline{A}$ is some constant. The variation of pressure field over a sphere of radius $\overline{A}$ is shown in the Figure~\ref{f:dipolepress}. Two lobes indicate the variation of pressure given by Equation~(\ref{eq:tpressuredi}). If we draw a radial vector with tail at the origin and its tip moving on the inner surface of the lobe, its magnitude gives the pressure at a point on the sphere in the direction of the vector.  We will find a plane of zero pressure which is parallel to the $y-z$ plane and passes through the origin. Total force acting on a circle in this plane is zero. 
 
\begin{figure}
\centering
\includegraphics[scale=0.4]{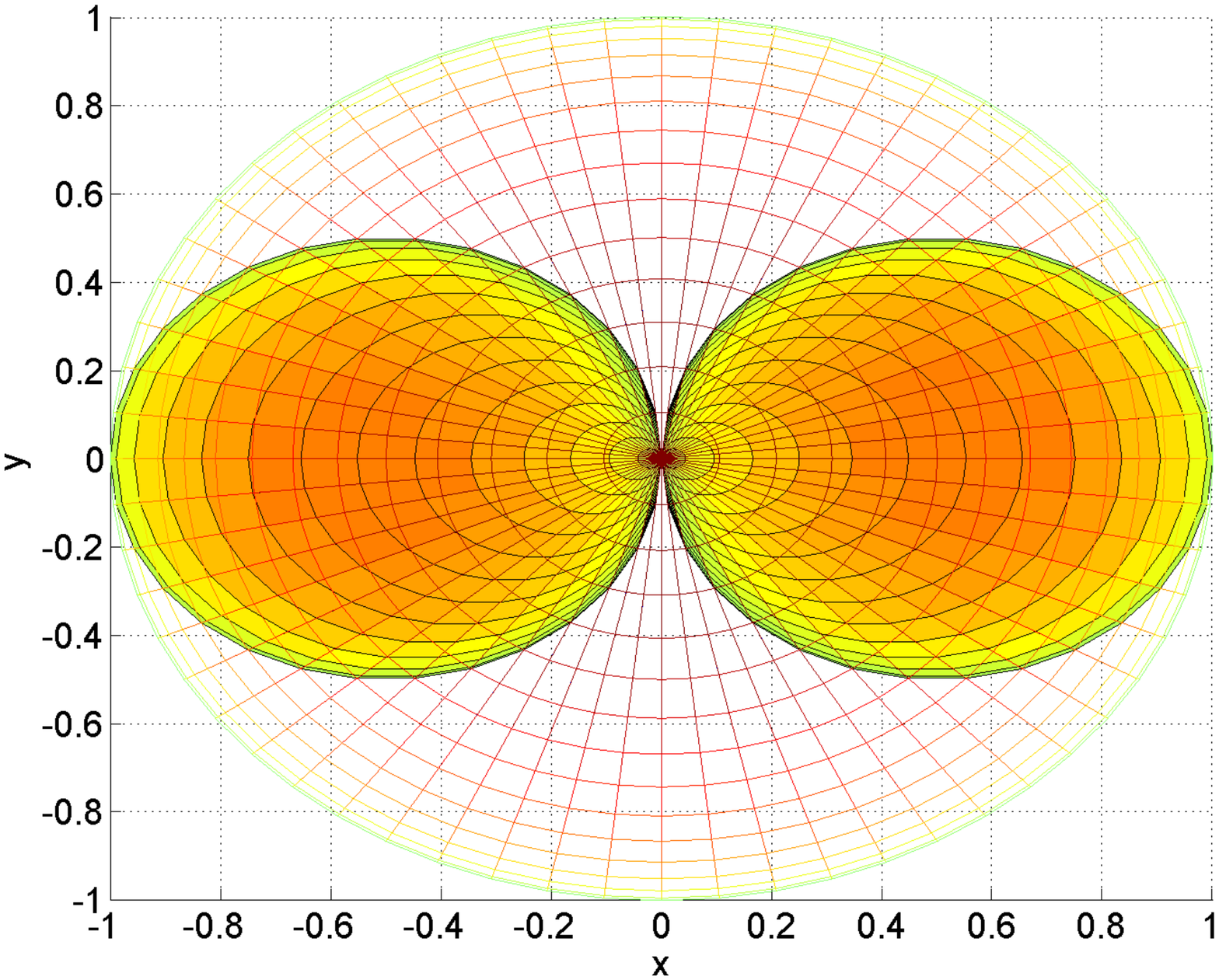}
\caption{Pressure contours on the $x-y$ plane of a dipole aligned with $x$ axis. $\overline{\textbf{p}}_d(\theta)=\overline{A} \cos\theta$; $\overline{A}=1$}
\label{f:dipolepressxyplane}
\end{figure}	
 
 Pressure field in a plane parallel to $x-y$ plane and passing through the origin is shown in Figure~\ref{f:dipolepressxyplane}. The expression for pressure field in this plane is given by

\begin{equation}
\label{eq:tpressuredip}
\overline{\textbf{p}}_d(\theta)=\overline{A} \cos\theta 
\end{equation}  

\begin{figure}
\begin{center}
\begin{tikzpicture}[scale=1][font=\small]
\draw[very thick,orange](2.5980,1.5)arc(30:60:3);
\draw[loosely dashed,thick](0,0)--(2.5980,1.5);
\draw[loosely dashed,thick](0,0)--(1.5,2.5980);
\draw[blue,thick,->](0,0)--(3.5,0);
\draw[blue](3.5,0)node[anchor=west]{x};
\draw[red!60!blue,->](0.5,0)arc(0:30:0.5);
\draw[blue](0.4330,0.18)node[anchor=west]{$\theta$};
\draw[red!60!blue,<->](0.6928,0.4)arc(30:60:0.8);
\draw[blue](0.6056,0.8556)node[anchor=west]{$d\theta$};
\draw[yellow!80!blue,thick,->](0,0)--(0,3.5);
\draw[blue](0,3.5)node[anchor=south]{y};
\draw[gray,thick,->](2.1213,2.1213)--(3.5355,3.5355);
\draw[blue](3.5355,3.5355)node[anchor=west]{$dF_{xy}$};
\draw[gray,thick,->](2.1213,2.1213)--(3.5355,2.1213);
\draw[blue](3.5355,2.1213)node[anchor=west]{$dF_x$};
\draw[gray,thick,->](2.1213,2.1213)--(2.1213,3.5355);
\draw[blue](2.1213,3.5355)node[anchor=south]{$dF_y$};
\draw[blue](1.2990,0.50)node[anchor=west]{$\overline{A}$};
\end{tikzpicture}
\end{center}
\end{figure}

Force acting on the differential element of the circle in the $x-y$ plane is

\begin{equation}
dF_{xy}=\overline{A}\cos\theta\,\,\overline{A}d\theta
\end{equation}

If we resolve the differential force along $x$ direction

\begin{equation}
dF_x=dF_{xy}\,\cos\theta
\end{equation}

Total force acting on the circle along the $x$ direction

\begin{equation}
F_x=\int_0^{2\pi} \overline{A}^2\cos^2\theta \,\,d\theta=\pi\overline{A}^2
\end{equation}

If we resolve the differential force along the $y$ direction

\begin{equation}
dF_y=dF_{xy}\,\sin\theta
\end{equation}

Total force acting on the circle along the $y$ direction

\begin{equation}
F_y=\int_0^{2\pi} \overline{A}^2\cos\theta\,\sin\theta \,\,d\theta=0
\end{equation}

Therefore total force acting on a circle in a plane parallel to $x-y$ plane passing through origin is along the $x$ direction and its magnitude is $\pi\overline{A}^2$.

\begin{figure}
\centering
\includegraphics[scale=0.4]{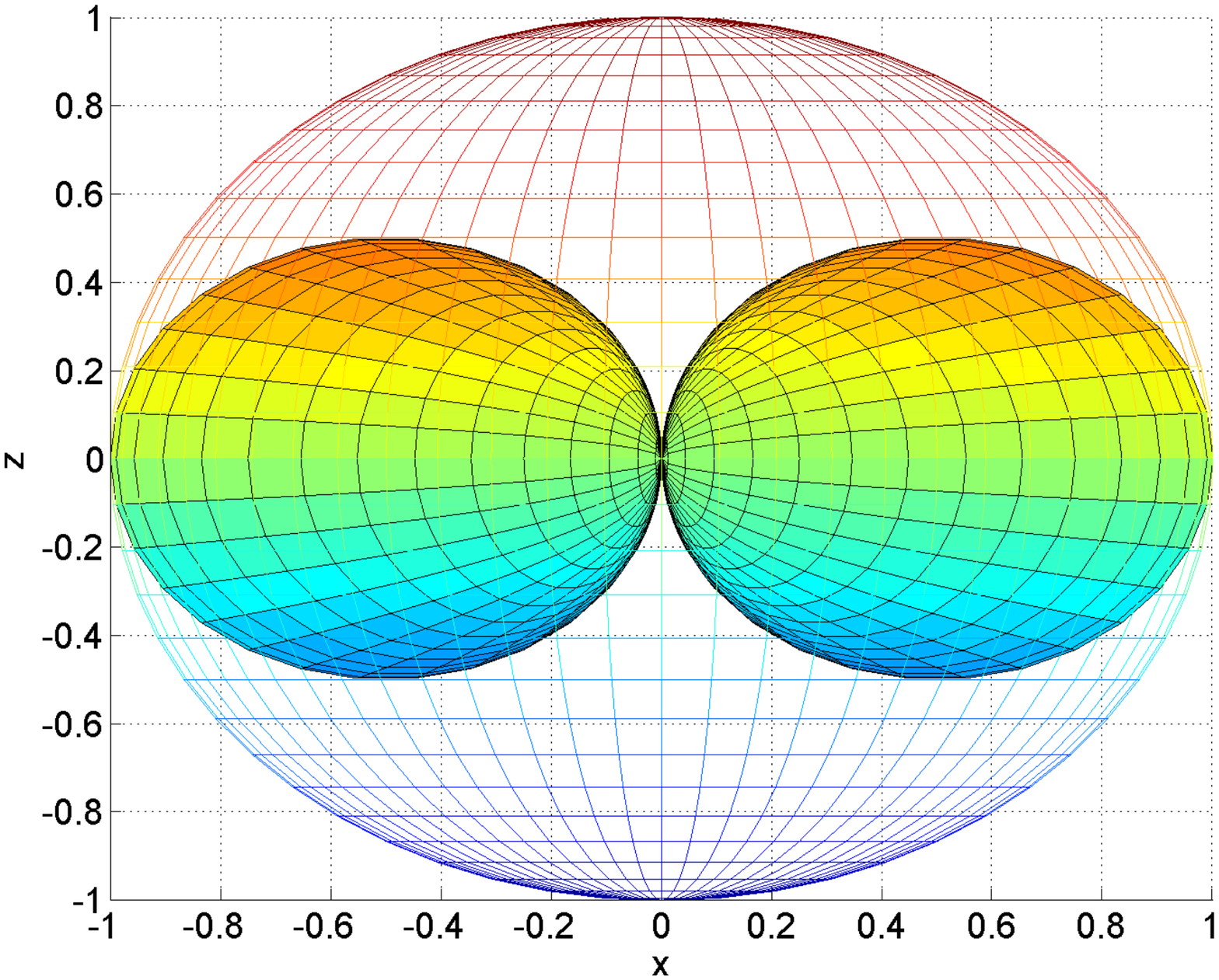}
\caption{Pressure contours on the $x-z$ plane of a dipole aligned with $x$ axis. $\overline{\textbf{p}}_d(\phi)=\overline{A} \sin\phi$; $\overline{A}=1$}
\label{f:dipolepressxzplane}
\end{figure}

Pressure field in a plane parallel to $x-z$ plane and passing through the origin is shown in Figure~\ref{f:dipolepressxzplane}. The expression for pressure field in this plane is given by

\begin{equation}
\label{eq:tpressuredip}
\overline{\textbf{p}}_d(\phi)=\overline{A} \sin\phi 
\end{equation}  

\begin{figure}
\begin{center}
\begin{tikzpicture}[scale=1][font=\small]
\draw[very thick,orange](2.5980,1.5)arc(30:60:3);
\draw[loosely dashed,thick](0,0)--(2.5980,1.5);
\draw[loosely dashed,thick](0,0)--(1.5,2.5980);
\draw[blue,thick,->](0,0)--(3.5,0);
\draw[blue](3.5,0)node[anchor=west]{x};
\draw[red!60!blue,->](0,0.5)arc(90:60:0.5);
\draw[blue](0.22,0.5030)node[anchor=south]{$\phi$};
\draw[red!60!blue,<->](0.6928,0.4)arc(30:60:0.8);
\draw[blue](0.6056,0.8556)node[anchor=west]{$d\phi$};
\draw[red,thick,->](0,0)--(0,3.5);
\draw[blue](0,3.5)node[anchor=south]{z};
\draw[gray,thick,->](2.1213,2.1213)--(3.5355,3.5355);
\draw[blue](3.5355,3.5355)node[anchor=west]{$dF_{xz}$};
\draw[gray,thick,->](2.1213,2.1213)--(3.5355,2.1213);
\draw[blue](3.5355,2.1213)node[anchor=west]{$dF_x$};
\draw[gray,thick,->](2.1213,2.1213)--(2.1213,3.5355);
\draw[blue](2.1213,3.5355)node[anchor=south]{$dF_z$};
\draw[blue](1.2990,0.50)node[anchor=west]{$\overline{A}$};
\end{tikzpicture}
\end{center}
\end{figure}

Force acting on the differential element of the circle in the $x-z$ plane is

\begin{equation}
dF_{xz}=\overline{A}\sin\phi\,\,\overline{A}d\phi
\end{equation}

If we resolve the differential force along the $x$ direction

\begin{equation}
dF_x=dF_{xz}\,\sin\phi
\end{equation}

Total force acting on the circle along the $x$ direction

\begin{equation}
F_x=\int_0^{2\pi} \overline{A}^2\sin^2\phi \,\,d\phi=\pi\overline{A}^2
\end{equation}

Similarly if we resolve the differential force along $z$ direction

\begin{equation}
dF_z=dF_{xz}\,\cos\phi
\end{equation}

Total force acting on the circle along the $z$ direction

\begin{equation}
F_z=\int_0^{2\pi} \overline{A}^2\,\sin\phi\,\cos\phi\,d\phi=0
\end{equation}

Therefore total force acting on a circle in a plane parallel to the $x-z$ plane passing through origin is along the $x$ direction only and its magnitude is $\pi\overline{A}^2$.

The assumptions that $\bar{k}_a\bar{l}<<1$ holds true, but in addition, the radius of the sphere $\bar{a}_s$ is such that $\bar{k}_a\bar{a}_s<<1$. This is automatically realized if we enforce the condition that $\bar{a}_s/\bar{l}<<1$. Note that the oscillations of the sphere or the rigid body that is vibrating should be of small amplitude. Only then can the dipole model be invoked to represent the pressure perturbations of an oscillating body. Given that these assumptions are satisfied, the dipole origin is placed at the equilibrium position of the vibrating body. The surface velocity of the dipole is the velocity of the vibrating mass.

\section{Results and Discussion}
\label{sec:resdis}

We first study the vibration behavior of the single degree of freedom oscillator with impact described in Sections~\ref{sec:vibmod} and \ref{sec:impmod}. The single degree of freedom oscillator is a linear system. Its mass impacting a barrier makes it nonlinear. That is, a monotone harmonic input to the system could result in a output that is aharmonic, or with many harmonic components, or with sub-harmonic or super-harmonic components. Impact with the barrier is modeled using a simple coefficient of restitution model. Vibration behavior, then, needs to be studied as a function of system parameters. The parameters that control the vibration response are linear natural frequency, damping ratio, excitation frequency, and excitation amplitude.  

\begin{figure}
\begin{center}
\includegraphics[width=\linewidth,height=9cm]{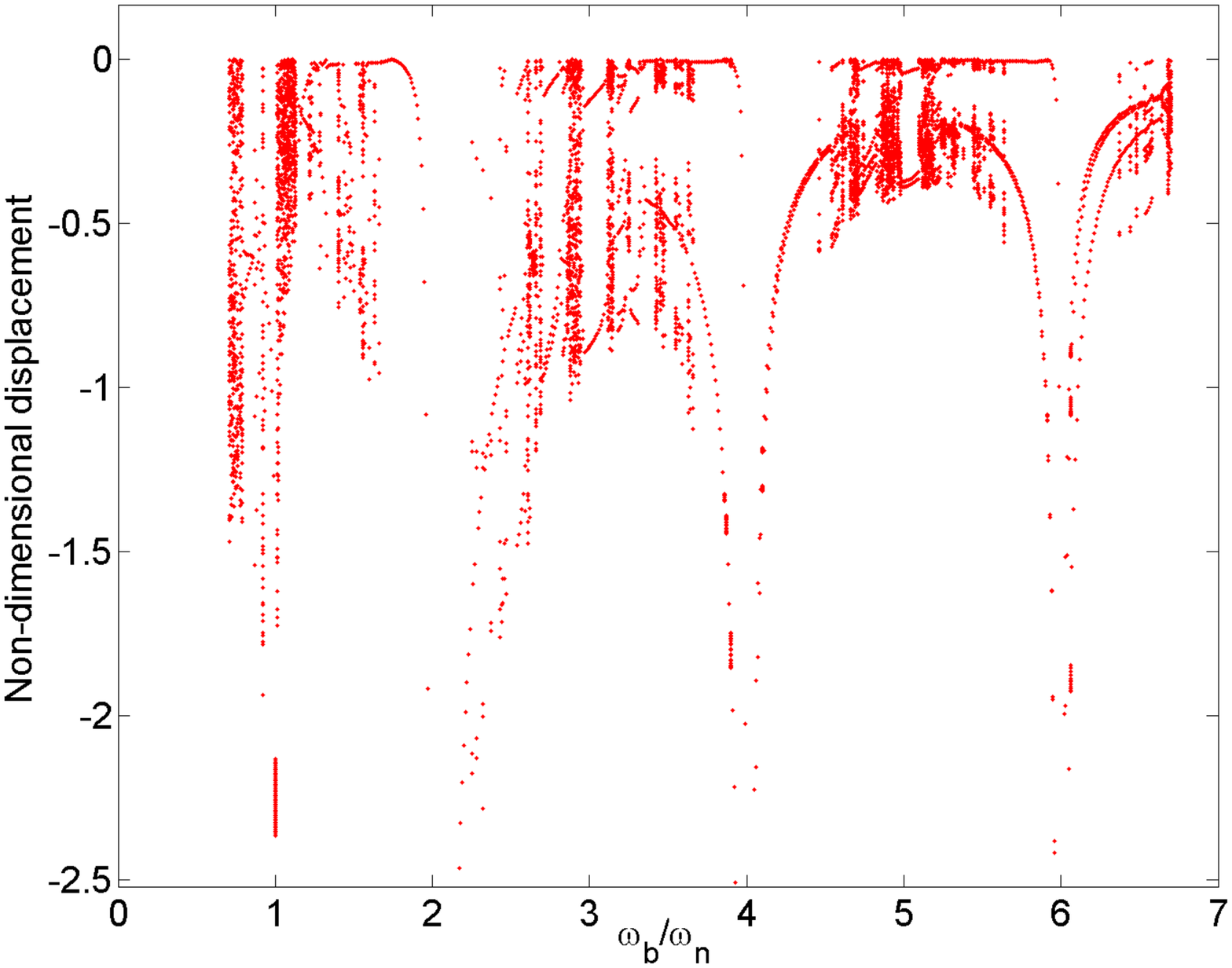}
 \caption{Bifurcation diagram: $e=1$, $\zeta=0.01$, $\overline{u}(0)=0$, $\overline{u}'(0)=0$,  $\overline{\delta}=0$.}
\label{f:bifurdia}
\end{center}
\end{figure}

For a single-degree-of-freedom nonlinear oscillator such as discussed in Section~\ref{sec:vibmod}, the bifurcation diagram is an useful tool for studying stability and response of the system to various system parameters. Shown in Figure~\ref{f:bifurdia} is the  non-dimensional displacement of the oscillator as a function of non-dimensionalized excitation frequency. The behavior of the system at various base excitation frequencies is determined by solving numerically Equation~(\ref{e:presen}) with the initial conditions $\overline{u}(0)=0$, $\overline{u}'(0)=0$. The procedure followed for plotting the bifurcation diagram is that for a given base excitation frequency $\overline{\omega}_b$, after $950$ impacts the least value of $\overline{u}(\tau)$ between any two impacts of the mass with the barrier is taken as the amplitude for that duration of motion. The amplitudes for the next $200$ impacts for a given $\overline{\omega}_b$ are evaluated  and plotted  in the bifurcation diagram. For instance, for frequencies close to $\overline{\omega}_b=2$, the mass bounces to the same amplitude for the $200$ impacts, so we see a single point in the bifurcation diagram. This implies that the response frequency is also periodic and the same period or frequency of the base excitation. For frequencies near $\overline{\omega}_b=2.9$, the mass bounces to different amplitudes, that leads to $200$ different points in the diagram. This implies that the response is not periodic or the motion does not repeat. One could, with some caution, say that the system shows possible chaotic behavior in the regions near $\overline{\omega}_b=2.9$.

In the discussion that follows, we will consider a reference system whose parameters corresponding to the vibro-impact system discussed in this chapter are $e=1$, $\zeta=0.01$, $\overline{\omega}_b=0.8$, $\overline{\delta}=0$, and $\overline{u}(0)=0,\overline{u}'(0)=0$. Variations to parameters such as $\omega_b$, $e$, and $\zeta$ about the reference values listed above and their influence on the vibration and acoustic response will be studied.

The acoustic model of the vibro-impact system has the following parameters: $\overline{c}=50$, $\overline{l}=0.001$, $\overline{a}=0.0001$. These are kept constant for all the simulations. The pressure is measured at the field point ($-100\,\overline{l}$, $100\,\overline{l}$). Note that this is a two-dimensional acoustic field.

\subsection{$e=1$, $\zeta=0.01$, $\overline{\omega}_b=0.8$, $\overline{\delta}=0$, $\overline{u}(0)=0$, $\overline{u}'(0)=0$.}
\label{sec:res1}

\begin{figure}
\centering
\includegraphics[width=0.8\linewidth,height=8cm]{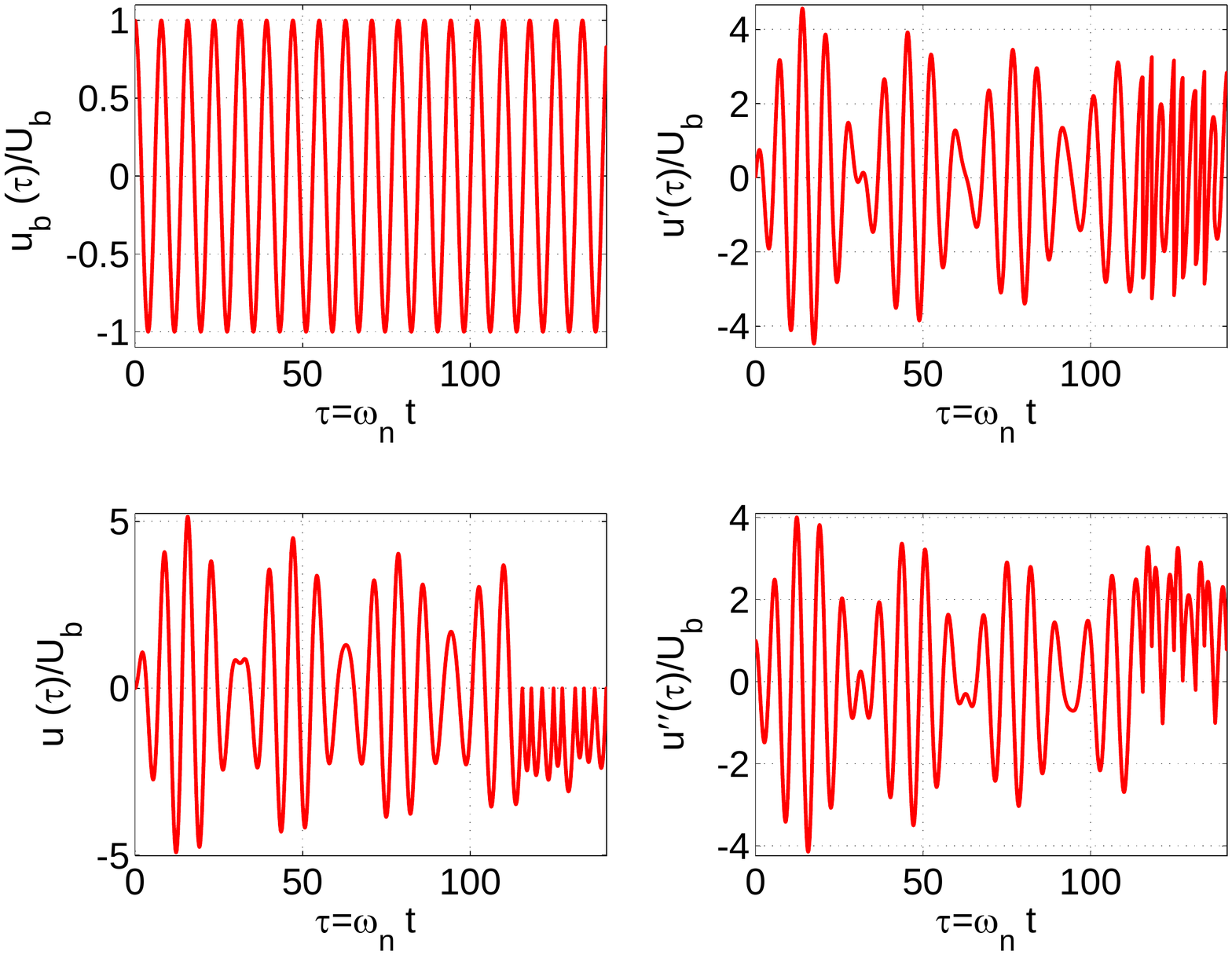}
 \caption{Transient response of the oscillator. $e=1$, $\overline{\omega}_b=0.8$, $\zeta=0.01$, $\overline{u}(0)=0$, $\overline{u}'(0)=0$,  $\overline{\delta}=0$.}
\label{f:bender vibration response1}
\end{figure}

\begin{figure}
\centering
\includegraphics[width=0.8\linewidth,height=8cm]{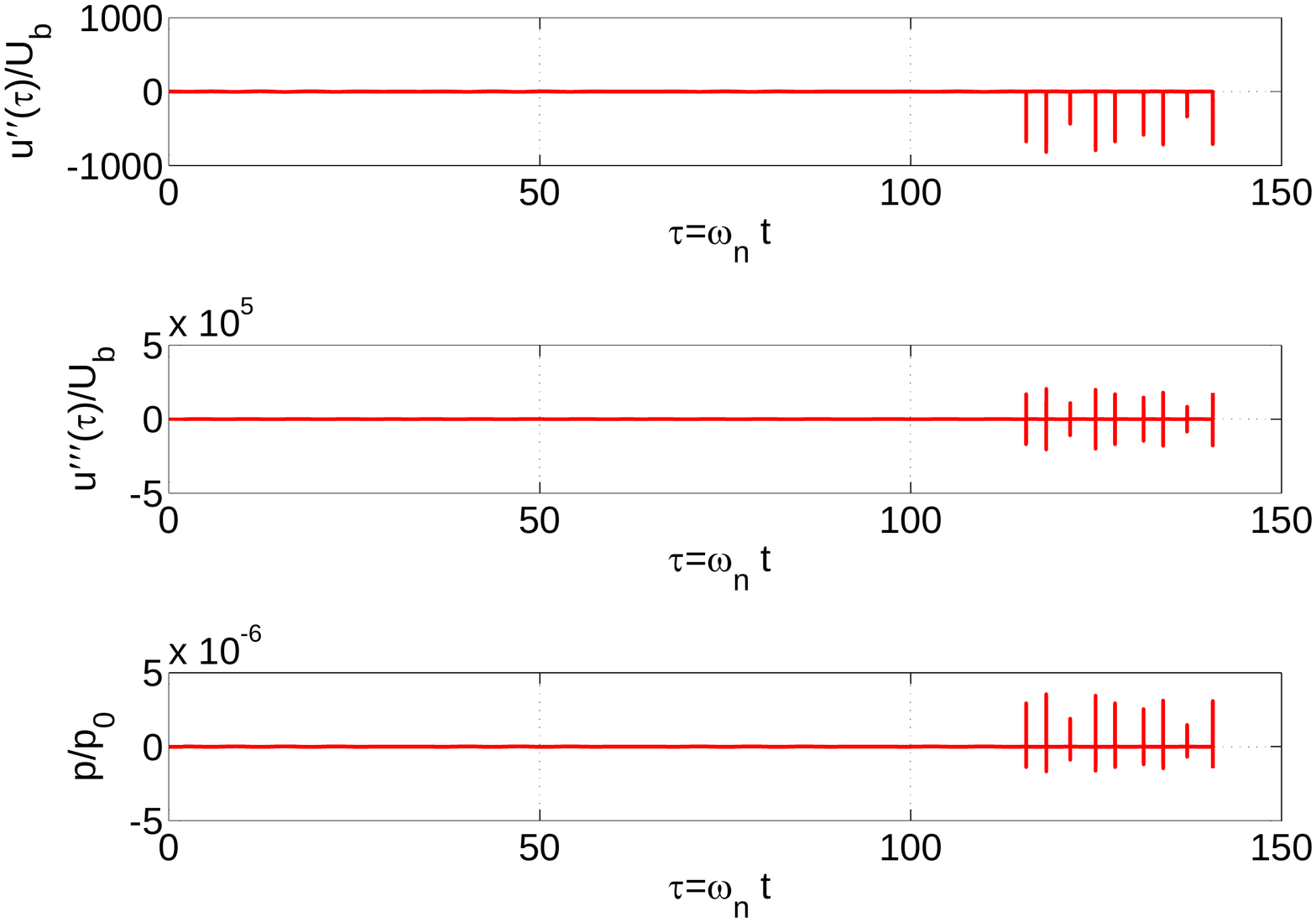}
 \caption{Transient pressure response at a point ($-100\,\overline{l}$, $100\,\overline{l}$). $\overline{c}=50$, $\overline{l}=0.001$, $\overline{a}=0.0001$,  $\overline{\rho}=10$, $e=1$, $\overline{\omega}_b=0.8$, $\zeta=0.01$, $\overline{u}(0)=0$, $\overline{u}'(0)=0$, $\overline{\delta}=0$.}
\label{f:bender pressure response1}
\end{figure}

We first determine the response of the reference system. For these system parameters, Equation~(\ref{e:presen}) is solved numerically. Transient response of the  system is shown in Figure{~\ref{f:bender vibration response1}}. Once the oscillator is about to complete its $16^{\text{th}}$ cycle of oscillation  about its equilibrium position, a massive barrier is placed at $\overline{u}(\tau)=0$.  The mass oscillates on one side of the equilibrium position once it starts impacting the barrier. The collision between the oscillator and the barrier is assumed to be elastic. From the velocity response we can observe that at the time of collision, the oscillator simply reverses its velocity with which it is impacting the barrier. The vibration response in this case too is random-like. This can be seen from the bifurcation diagram Figure{~\ref{f:bifurdia}} for the base excitation frequency value $\overline{\omega}_b=0.8$.
 
Pressure response at a point in the free field in the $xy$ plane is calculated using Equation~(\ref{eq:dipole pressure in xy coordinate system}) and is shown in Figure{~\ref{f:bender pressure response1}}. It is assumed that the barrier that is placed at the equilibrium position of the mass, is anechoic and its presence does not affect the acoustic pressure field as it simulates the acoustic free field condition.  From Equation~(\ref{eq:dipole pressure in xy coordinate system}), we can say that whatever the pressure difference experienced at time $\tau=0$ at point ($\bar{x}_1(\tau),\bar{y}_1(\tau))$ is experienced at point ($\bar{x}_2(\tau),\bar{y}_2(\tau))$ after time $\tau=\frac{\bar{r}}{\bar{c}}$, where $\bar{r}=\sqrt{(\bar{x}_2(\tau)-\bar{x}_1(\tau))^2+(\bar{y}_2(\tau)-\bar{y}_1(\tau))^2}$. Thus there is a time lag between acceleration and jerk to that of pressure response. 

As shown in Figure{~\ref{f:bender pressure response1}}, during impact, acceleration and jerk resulting from the impact force are added to the acceleration and jerk of the vibrating mass to obtain the total acceleration and jerk response. We can observe from the first plot in Figure{~\ref{f:bender pressure response1}} that acceleration and jerk resulting from the impact force are quite high as compared to acceleration and jerk of the vibrating mass as shown in Figure~\ref{f:bender vibration response1}. This is because the plot for acceleration in Figure{~\ref{f:bender pressure response1}} contains the acceleration due to the impact computed using Equation~(\ref{eq:impact force acceleration amplitude}). Since the acoustic pressure depends on acceleration and jerk response, its magnitude is very high at the time of impact as compared to the pressure field generated by the vibrating mass.   
 
 \subsection{$e=1$, $\zeta=0.01$, $\overline{\omega}_b=2.9$, $\overline{\delta}=0$, $\overline{u}(0)=0$, $\overline{u}'(0)=0$.}
\label{sec:res2}

\begin{figure}
\centering
\includegraphics[width=0.8\linewidth,height=8cm]{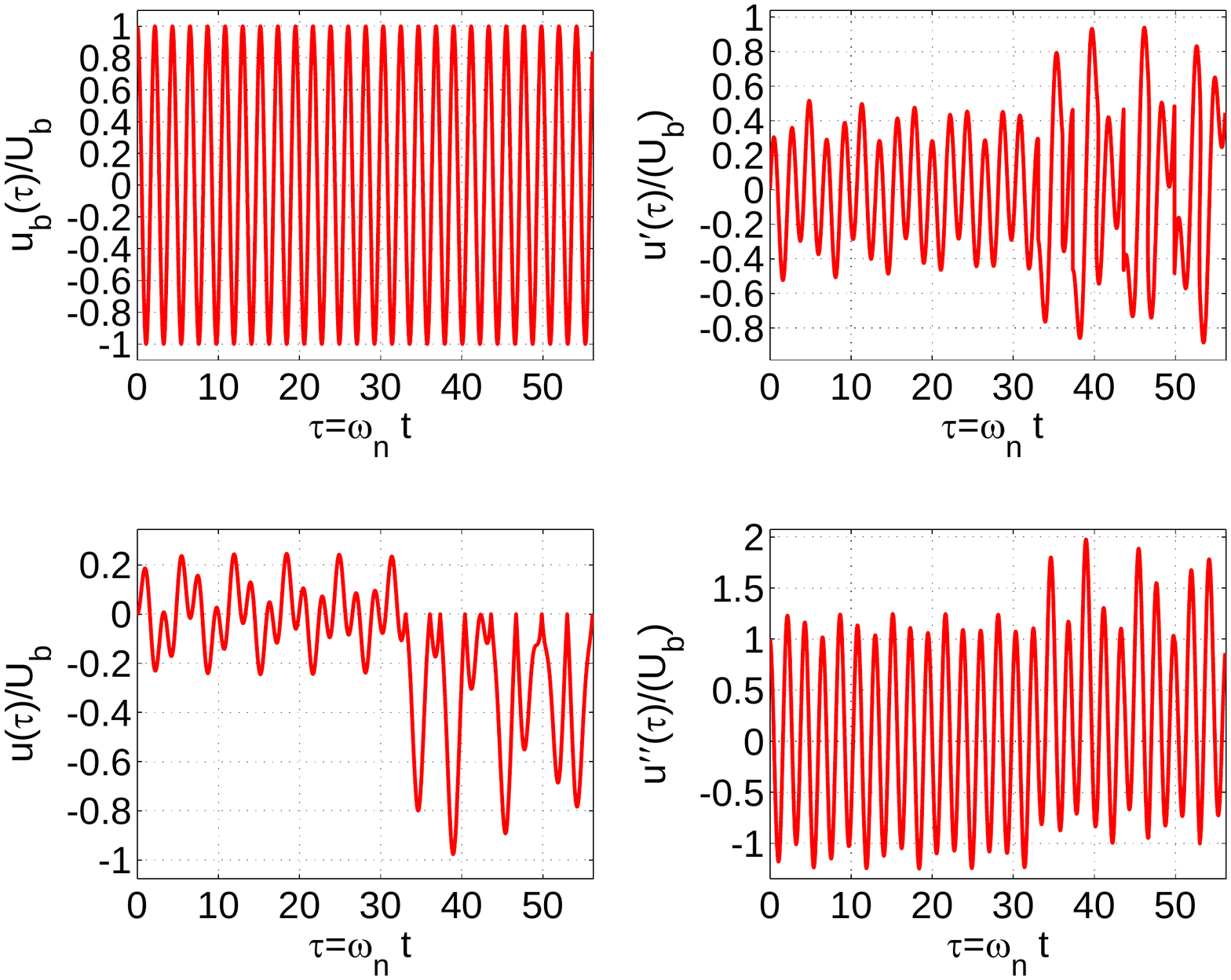}
 \caption{Transient response of the oscillator: $e=1$, $\overline{\omega}_b=2.9$, $\zeta=0.01$, $\overline{u}(0)=0$, $\overline{u}'(0)=0$,  $\overline{\delta}=0$.}
\label{f:bender vibration response2}
\end{figure}

\begin{figure}
\centering
\includegraphics[width=0.8\linewidth,height=8cm]{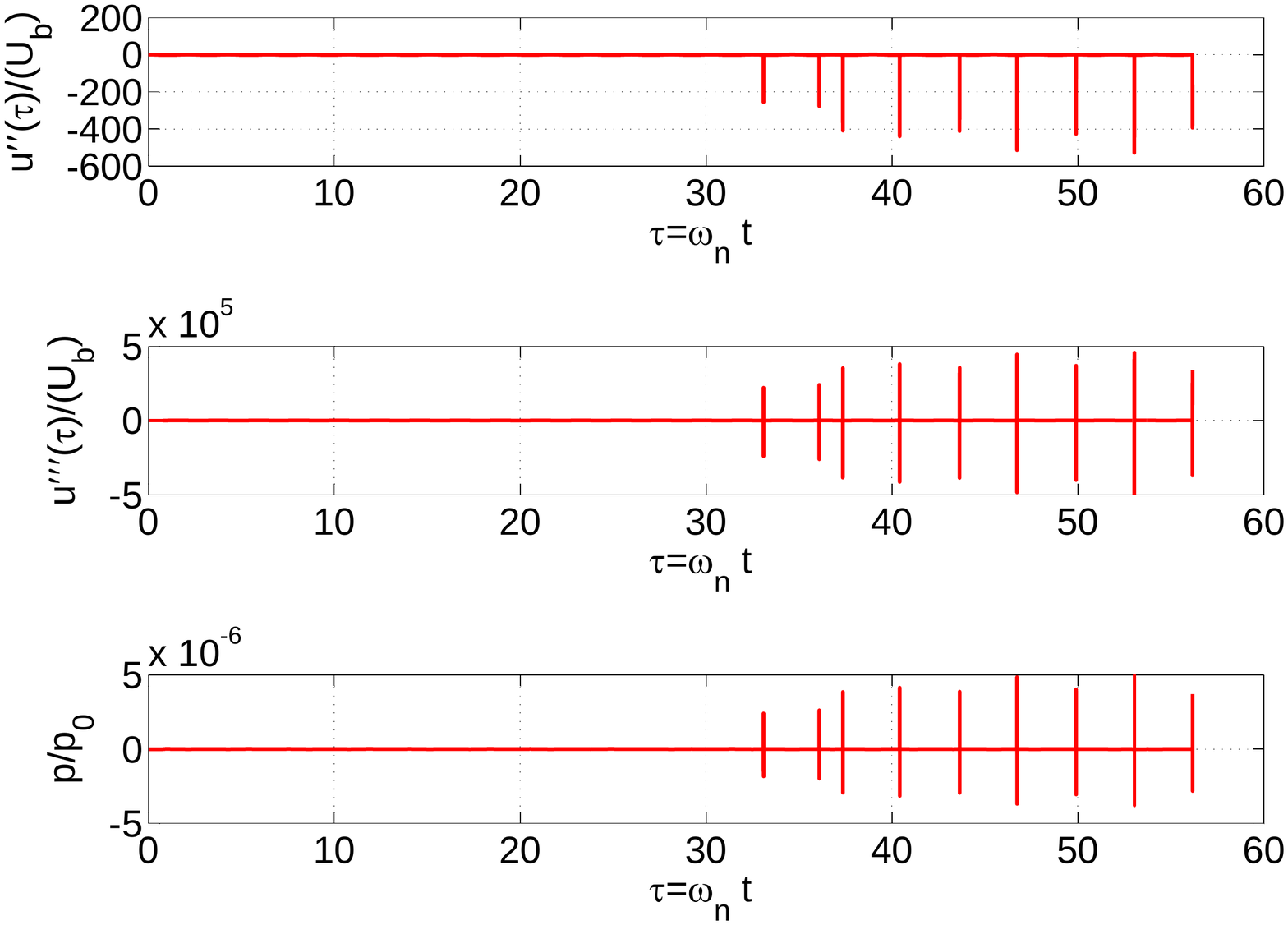}
 \caption{Transient pressure response at a point ($-100\,\overline{l}$, $100\,\overline{l}$). $\overline{c}=50$, $\overline{l}=0.001$, $\overline{a}=0.0001$,  $\overline{\rho}=10$, $e=1$, $\overline{\omega}_b=2.9$, $\zeta=0.01$, $\overline{u}(0)=0$, $\overline{u}'(0)=0$,  $\overline{\delta}=0$.}
\label{f:bender pressure response2}
\end{figure}

Note that in this set of system parameters, the base excitation frequency is $\overline{\omega}_b=2.9$. The transient response of the oscillator is shown in Figure{~\ref{f:bender vibration response2}}. Note that with reference to the bifurcation diagram shown in Figure{~\ref{f:bifurdia}} that the system shows chaotic-like behavior. Comparing the vibration response to the case $\overline{\omega}_b=0.8$, the displacement amplitude is higher.

Transient acoustic pressure response is shown in the Figure{~\ref{f:bender pressure response2}}. From Equation~(\ref{eq:impact force acceleration amplitude}), we can see that the amplitude of acceleration $\overline{A}$ resulting from the impact force is directly proportional to base excitation frequency $\overline{\omega}_b$. So as we increase the base excitation frequency, the amplitudes of acceleration and jerk resulting from the impact force will increase and as a result of that  magnitude of pressure response is high as compared to the pressure response as shown in Figure{~\ref{f:bender pressure response1}}.

\subsection{$e=1$, $\zeta=0.1$, $\overline{\omega}_b=0.8$, $\overline{\delta}=0$, $\overline{u}(0)=0$, $\overline{u}'(0)=0$.}
\label{sec:res3}

\begin{figure}
\centering
\includegraphics[width=0.8\linewidth,height=8cm]{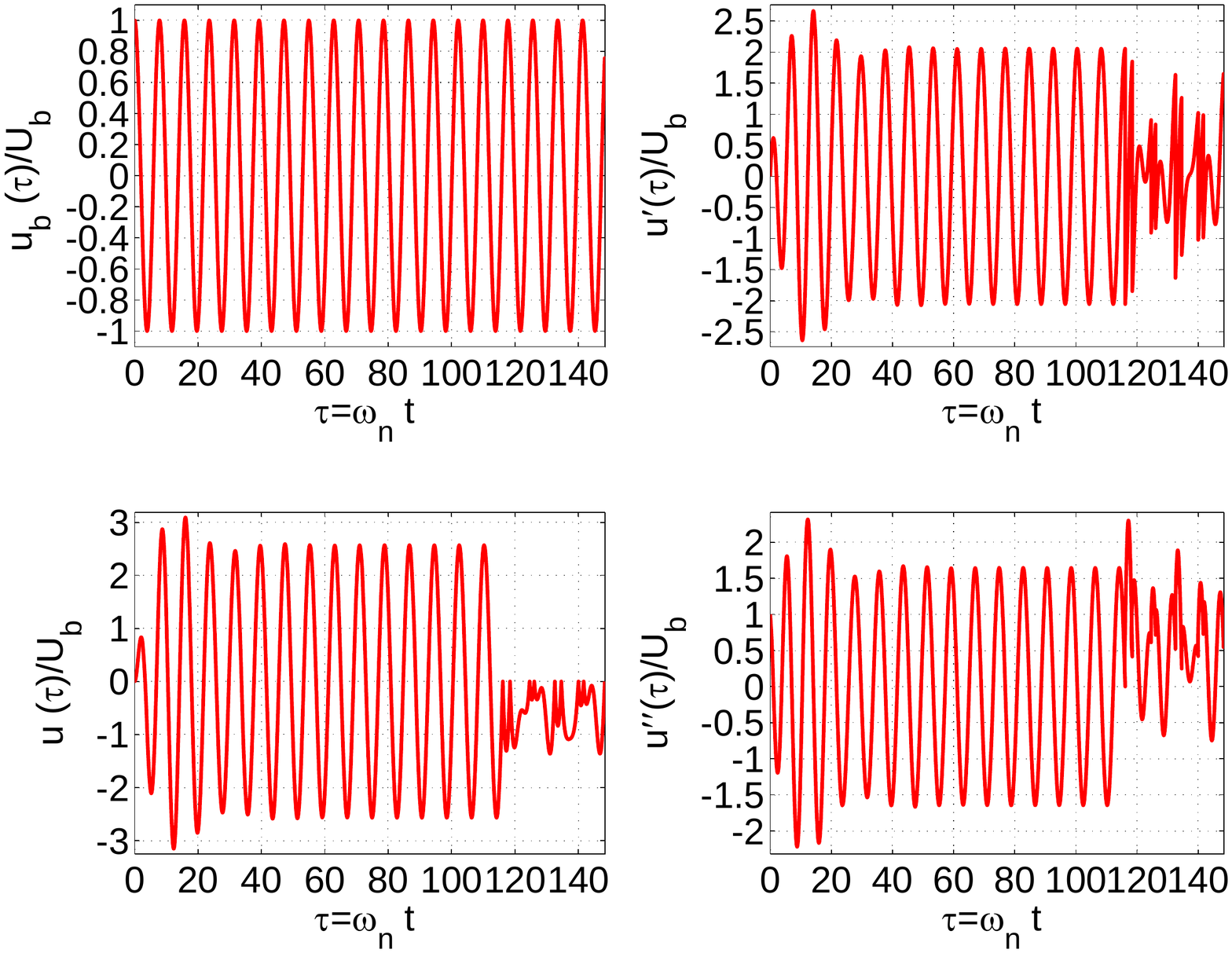}
\caption{Transient response of the oscillator: $e=1$, $\overline{\omega}_b=0.8$, $\zeta=0.1$, $\overline{u}(0)=0$, $\overline{u}'(0)=0$,  $\overline{\delta}=0$.}
\label{f:bender vibration response3}
\end{figure}

\begin{figure}
\centering
\includegraphics[width=0.8\linewidth,height=8cm]{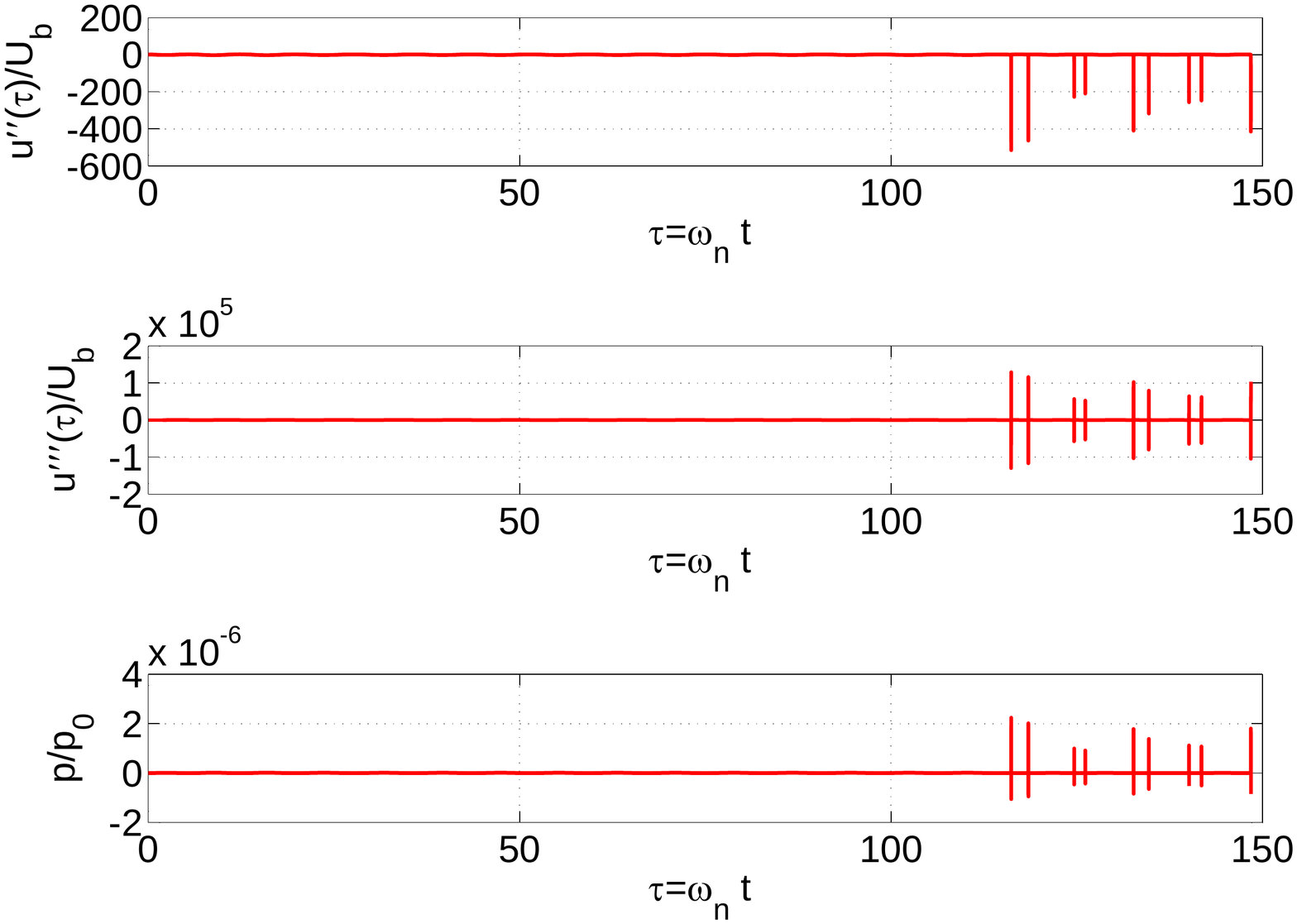}
\caption{Transient pressure response at a point ($-100\,\overline{l}$, $100\,\overline{l}$), $\overline{c}=50$, $\overline{l}=0.001$, $\overline{a}=0.0001$,  $\overline{\rho}=10$, $e=1$, $\overline{\omega}_b=0.8$, $\zeta=0.1$, $\overline{u}(0)=0$, $\overline{u}'(0)=0$,  $\overline{\delta}=0$.}
\label{f:bender pressure response3}
\end{figure}

In this case, the damping is increased by a factor of $10$ from that in the reference system Section~\ref{sec:res1}. The transient vibration response is shown in Figure~\ref{f:bender vibration response3}. Note that the increase in damping lowers not only the displacement response but also the velocity response of the oscillator.

Transient pressure response is shown in Figure{~\ref{f:bender pressure response3}}. From Equation~(\ref{eq:impact force acceleration amplitude}), we can see that the  amplitude of acceleration $\overline{A}$ resulting from the impact force is directly proportional to the velocity $\overline{u}{'}(\tau^{\ast}_{-})$ with which the mass is impacting the barrier. From the velocity response in Figure~\ref{f:bender vibration response3} we can see that the impact velocity $\overline{u}{'}(\tau^{\ast}_{-})$  is less compared to the impact velocity as shown in the Figure{~\ref{f:bender vibration response1}}. So by increasing the damping ratio, there is  a decrement in the amplitude of  acceleration and jerk  resulting from impact force which will in turn result in less acoustic pressure as compared the pressure field shown in the Figure{~\ref{f:bender pressure response1}}.

\subsection{$e=1$, $\zeta=0.01$, $\overline{\omega}_b=0.8$, $\overline{\delta}=0.1$, $\overline{u}(0)=0$, $\overline{u}'(0)=0$.}
\label{sec:res4}

\begin{figure}
\centering
\includegraphics[width=0.8\linewidth,height=8cm]{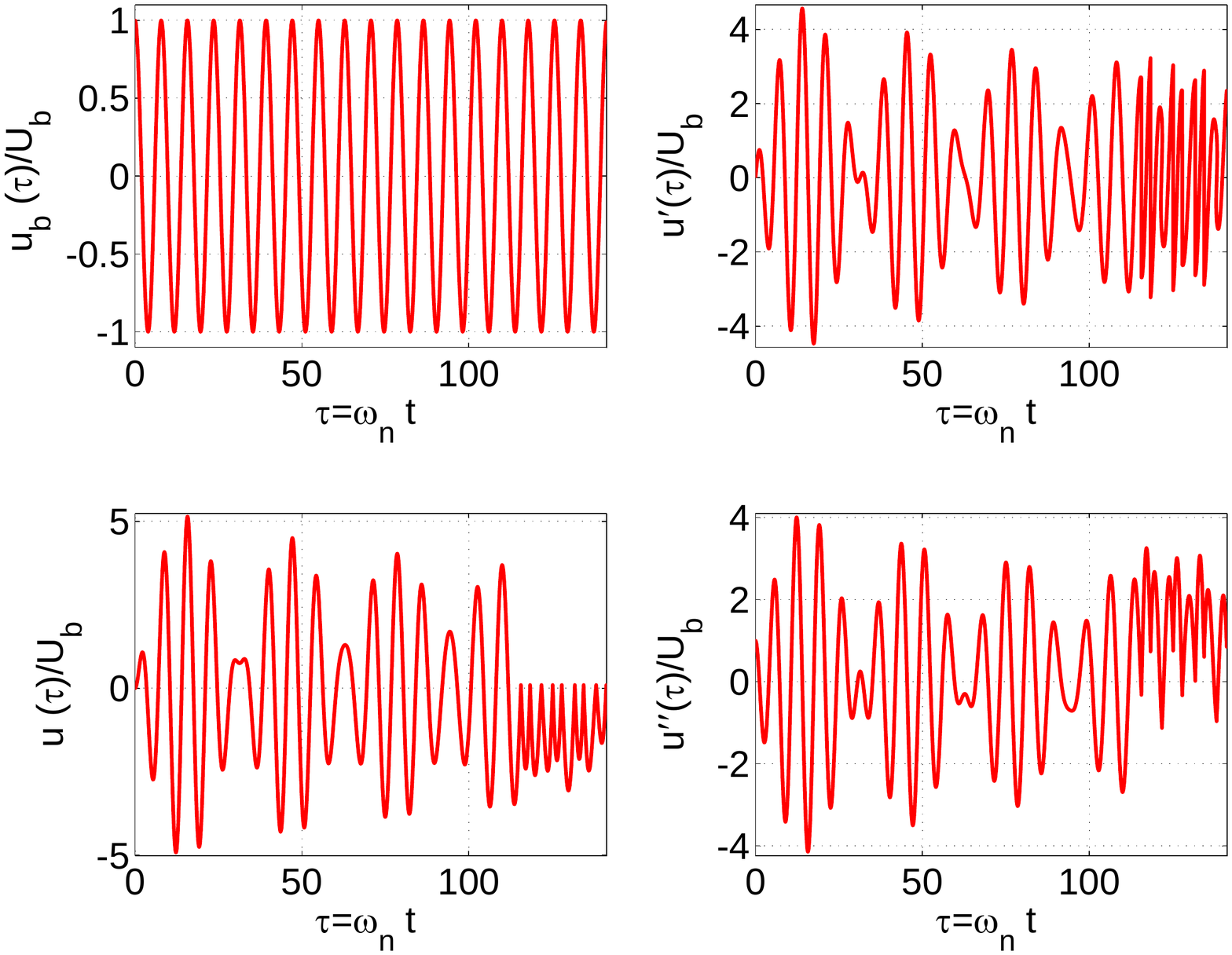}
\caption{Transient response of the oscillator: $e=1$, $\overline{\omega}_b=0.8$, $\zeta=0.01$, $\overline{u}(0)=0$, $\overline{u}'(0)=0$,  $\overline{\delta}=0.1$.}
\label{f:bender vibration response4}
\end{figure}

\begin{figure}
\centering
\includegraphics[width=0.8\linewidth,height=8cm]{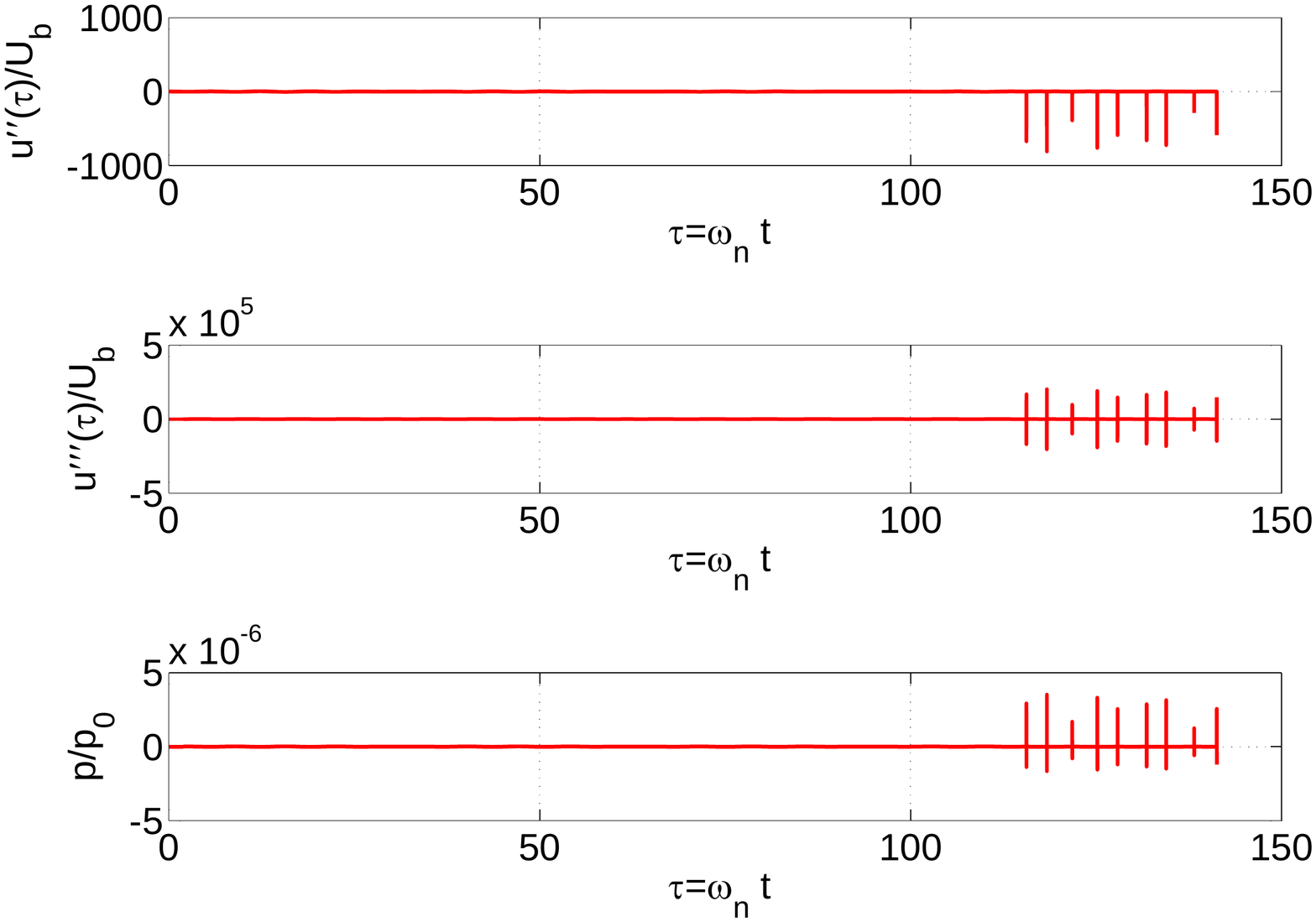}
\caption{Transient pressure response at a point ($-100\,\overline{l}$, $100\,\overline{l}$), $\overline{c}=50$, $\overline{l}=0.001$, $\overline{a}=0.0001$,  $\overline{\rho}=10$, $e=1$, $\overline{\omega}_b=0.8$, $\zeta=0.01$, $\overline{u}(0)=0$, $\overline{u}'(0)=0$,  $\overline{\delta}=0.1$.}
\label{f:bender pressure response4}
\end{figure}

In this case the distance where the barrier is placed $\overline{\delta}$ is placed is shifted to the right of the equilibrium position of the oscillator mass. Now $\overline{\delta}=0.1$. Transient response of the oscillator corresponding these system parameters is shown in the Figure{~\ref{f:bender vibration  response4}}. 

Transient pressure response is shown in Figure{~\ref{f:bender pressure response4}}. By placing the barrier at certain distance away from the equilibrium position, we can see from Figure{~\ref{f:bender vibration response4}} that there is not much difference in the impact velocity $\overline{u}{'}(\tau^{\ast}_{-})$  as compared to the one shown in Figure{~\ref{f:bender vibration response1}}. So there is no significant rise in the magnitude of acoustic pressure as compared to the pressure field shown in Figure{~\ref{f:bender pressure response1}}.

\subsection{Influence of coefficient of restitution}
\label{sec:infcoeres}

\begin{figure}
\centering
\includegraphics[width=0.8\linewidth,height=8cm]{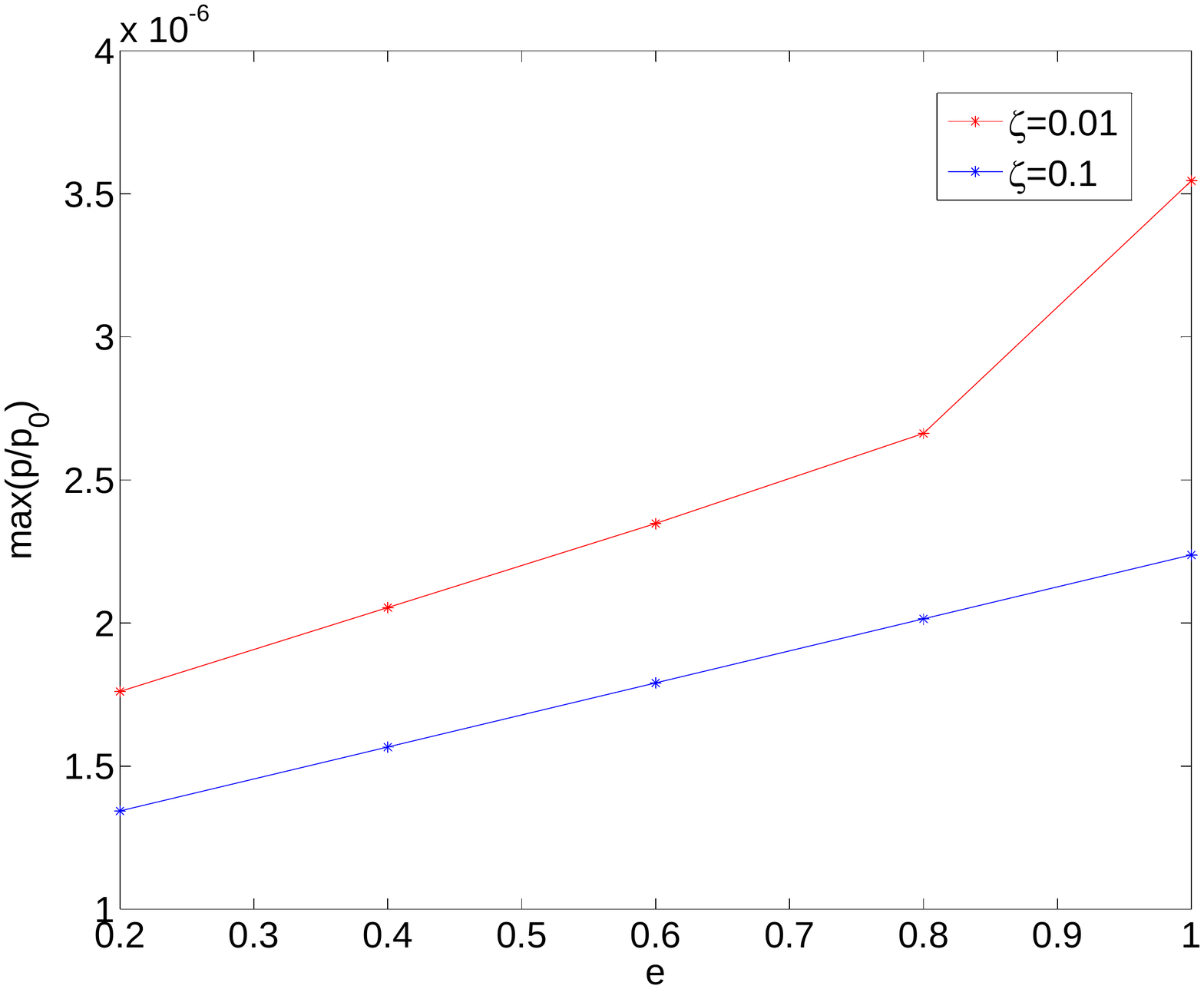}
\caption{Maximum acoustic pressure ratio as function of coefficient of restitution. ($-100\,\overline{l}$, $100\,\overline{l}$), $\overline{c}=50$, $\overline{l}=0.001$, $\overline{a}=0.0001$,  $\overline{\rho}=10$, $\overline{\omega}_b=0.8$, $\overline{u}(0)=0$, $\overline{u}'(0)=0$,  $\overline{\delta}=0.1$.}
\label{f:maxpvse}
\end{figure}

For the reference system as described in Section~\ref{sec:res1}, the coefficient of restitution is varied from the reference value of $e=1.0$. As the coefficient of restitution decreases, it implies that momentum and energy is lost in the collision with the barrier. Figure~\ref{f:maxpvse} shows the variation of the maximum value of the pressure ratio $\frac{p}{p_0}$ as a function of $e$ for two values of damping ratio $\zeta$. As the coefficient of restitution decreases, the maximum acoustic pressure radiated due to vibration and impact also reduces. This is reasonable as momentum and energy of the oscillator is lost as the collision becomes more lossy.

\section{Conclusion}
\label{sec:con}

In this paper, we have derived the relations needed to model sound generated due to vibration and impact. We have only considered the sound field in free space. We have considered a single-degree-of-freedom oscillator whose motion is  constrained  by a barrier which is introduced after the oscillator made certain number of oscillations about its equilibrium position. The acoustic pressure resulting from the impact force is high as compared to the pressure field resulting from the vibrating mass. It is the acceleration amplitude resulting from the impact force which mainly affects the peak sound pressure. By varying the system parameters, the response of the oscillator changes considerably which in turn changes the acceleration amplitude of the impact force and effects the peak sound pressure. The coefficient of restitution and damping ratio are the significant parameters that influence the acoustic pressure field generated by the vibro-impact oscillator.

\bibliographystyle{unsrt}
\bibliography{nspkv}

\begin{thebibliography}{10}

\bibitem{Akay1983}
A.~Akay and M.T Bengisu.
\newblock Transient acoustic radiation from impacted beam-like structures.
\newblock {\em Journal of Sound and Vibration}, 91:135--145, 1983.

\bibitem{Akay1978}
A.~Akay and T.H. Hodgson.
\newblock Acoustic radiation from the elastic impact of a sphere with a slab.
\newblock {\em Applied Acoustics}, 11:285--303, 1978.

\bibitem{Akay1983a}
A.~Akay and M.~Latcha.
\newblock Sound radiation from an impact-excited clamped circular plate in an
  infinite baffle.
\newblock {\em Journal of the Acoustical Society of America}, 74:640--648,
  1983.

\bibitem{Koss1973}
L.L. Koss and R.J. Alfredson.
\newblock Transient sound radiation by spheres undergoing an elastic collision.
\newblock {\em Journal of Sound and Vibration}, 27:59--75, 1973.

\bibitem{Koss1974}
L.L. Koss.
\newblock Transient sound from colliding spheres-inelastic collision.
\newblock {\em Journal of Sound and Vibration.}, 36:555--562, 1974.

\bibitem{Yufang1992}
W.~Yufang and T.~Zhongfang.
\newblock Sound radiated form the impact of two cylinders.
\newblock {\em Journal of Sound and Vibration}, 152:295--303, 1992.

\bibitem{Walker1996}
J.S. Walker and T~Soule.
\newblock Chaos in a simple impact oscillator:the bender bouncer.
\newblock {\em American Journal of Physics}, 64:397--409, 1996.
\newblock Seminare Maurey-Schwartz (1975-1976).

\bibitem{Tufillaro1986}
N.~B. Tufillaro and A.~M. Albano.
\newblock Chaotic dynamics of a bouncing ball.
\newblock {\em American Journal of Physics}, 54:939--944, 1986.

\bibitem{Troccaz2000}
P.~Troccaz, R.~Woodcook, and F.~Laville.
\newblock Acoustic radiation due to the inelastic impact of a sphere on a
  rectangular plate.
\newblock {\em Journal of the Acoustical Society of America}, 108:2197--2202,
  2000.

\bibitem{Fahy2000}
F.J. Fahy.
\newblock {\em Foundations of Engineering Acoustics}.
\newblock Academic Press, 2000.

\bibitem{Kinsler2000}
L.~E. Kinsler, A.~R. Frey, A.~B. Coppens, and J.~V. Sanders.
\newblock {\em Fundamentals of Acoustics}.
\newblock John Wiley and Sons, Inc., New York, 2000.

\end{thebibliography}

\end{document}